\documentclass[conference]{IEEEtran}
\ifCLASSINFOpdf
\else
\fi
\usepackage{balance}  
\usepackage{graphics} 
\usepackage{txfonts}
\usepackage{times}    
\usepackage[pdftex]{hyperref}
\usepackage{color}
\usepackage{textcomp}
\usepackage{booktabs}
\usepackage{ccicons}

\usepackage{url}
\usepackage{fancybox}
\usepackage{multirow}
\usepackage{flushend}
\usepackage{booktabs}
\usepackage{tabularx}
\usepackage{comment}
\usepackage{array}
\usepackage[flushleft]{threeparttable}
\usepackage{mdframed}
\graphicspath{{}{images/}{dia/}}
\DeclareGraphicsExtensions{.pdf,.png}

\usepackage{listings}
\usepackage{courier}
\usepackage{hyperref}

\usepackage{xpatch}

\xpatchcmd{\refstepcounter}{%
  \stepcounter{#1}%
}{%
  \stepcounter{#1}%
  \xdef\scr{\number\value{#1}}%
}{\typeout{success}}{\typeout{failure}}

\newcounter{o}
\usepackage{soul}
\usepackage{tikz}
\definecolor{1c1}{RGB}{188,162,6}
\definecolor{1c2}{RGB}{137,129,80}
\definecolor{1c3}{RGB}{239,167,31}
\definecolor{1c4}{RGB}{88,194,241}
\definecolor{1c5}{RGB}{6,180,188}

\tikzset{mynode/.style={draw=white,solid,circle,fill=green,inner sep=1pt, thick,
text=black}}
\tikzset{arrow line/.style={dashed, line width= 2.5pt, color=#1}}

\def\bf{\textbf}

\def\fig {Figure~}

\def\sec {Section~}

\def\it{\textit}

\def\tt{\mct}

\newcommand{\mct}[1]{{\footnotesize {\texttt {#1}}}}

\usepackage{paralist}

\newcommand{\nd}{\vspace{1mm}\noindent}
\usepackage[small,bf]{caption}
\usepackage{tikz}

\lstdefinestyle{inlinecode}{basicstyle={\ttfamily\scriptsize\bfseries}}

\newcommand{\urls}[1]{{\scriptsize\url{#1}}}
\usepackage{tcolorbox}

\newcommand{\dq}[1]{\href{https://stackoverflow.com/questions/#1/}{$Q_{#1}$}}

\usepackage{paralist}
\usepackage[outercaption]{sidecap}
\usepackage [autostyle, english = american]{csquotes}
\MakeOuterQuote{"}
\newcounter{scn}
\usepackage[shortlabels]{enumitem}
\usepackage{tikz}
\usepackage{styles/pgf-pie}
\usetikzlibrary{positioning,shadows}

\newif\ifpienumberinlegend
\pgfkeys{/number in legend/.code=
    \expandafter\let\expandafter\ifpienumberinlegend
    \csname if#1\endcsname
    \ifpienumberinlegend

    \def\beforenumber##1\afternumber{}%
    \fi,
    /number in legend/.default=true
}
\definecolor{1c1}{RGB}{188,162,6}
\definecolor{1c2}{RGB}{137,129,80}
\definecolor{1c3}{RGB}{239,167,31}
\definecolor{1c4}{RGB}{88,194,241}
\definecolor{1c5}{RGB}{6,180,188}

\tikzset{mynode/.style={draw=white,solid,circle,fill=green,inner sep=1pt, thick,
text=black}}
\tikzset{arrow line/.style={dashed, line width= 2.5pt, color=#1}}

\usepackage{bchart}

\usepackage{xcolor}
\usepackage[numbers]{natbib}
\definecolor{ao(english)}{rgb}{0.0, 0.5, 0.0}

\def\test#1{%
    \ifnum #1 > 0
      #1
    \fi
}

\newcommand{\sixbars}[6]{
{{\color{black}\rule{#1pt}{4pt}} \test{#1}}
{{\color{ao(english)}\rule{#2pt}{4pt}} \test{#2}}
{{\color{magenta}\rule{#3pt}{4pt}} \test{#3}}
{{\color{red}\rule{#4pt}{4pt}} \test{#4}}
{{\color{cyan}\rule{#5pt}{4pt}} \test{#5}}
{{\color{orange}\rule{#6pt}{4pt}} \test{#6}}
}

\usepackage{balance}

\begin{document}
\title{An Empirical Study of Developer Discussions on Low-Code Software Development Challenges}
\author{\IEEEauthorblockN{Md Abdullah Al Alamin\IEEEauthorrefmark{1}, Sanjay Malakar\IEEEauthorrefmark{2}, Gias Uddin\IEEEauthorrefmark{1}, Sadia Afroz\IEEEauthorrefmark{2}, Tameem Bin Haider\IEEEauthorrefmark{2}, Anindya Iqbal\IEEEauthorrefmark{2}}
\IEEEauthorblockA{\IEEEauthorrefmark{1}University of Calgary, \IEEEauthorrefmark{2}Bangladesh University of Engineering and Technology
%
}}

\IEEEtitleabstractindextext{%
\begin{abstract}

Low-code software development (LCSD) is an emerging paradigm that combines
minimal source code with interactive graphical interfaces to promote rapid
application development. LCSD aims to democratize application development to
software practitioners with diverse backgrounds. Given that LCSD is relatively a new
paradigm, it is vital to learn about the challenges developers face during
their adoption of LCSD platforms. The online developer forum, Stack Overflow
(SO), is popular among software developers to ask for solutions to their technical
problems. We observe a growing body of posts in SO with discussions of LCSD
platforms. In this paper, we present an empirical study of around 5K SO posts
(questions + accepted answers) that contain discussions of nine popular LCSD
platforms. We apply topic modeling on the posts to determine the types of topics
discussed. We find 13 topics related to LCSD in SO. The 13 topics are grouped
into four categories: Customization, Platform Adoption, Database Management, and
Third-Party Integration. More than 40\% of the questions are about
customization, i.e., developers frequently face challenges with customizing user interfaces or services offered by LCSD platforms. The
topic ``Dynamic Event Handling'' under the ``Customization'' category is the most
popular (in terms of average view counts per question of the topic) as well as
the most difficult. It means that developers frequently search for customization solutions such as how to attach dynamic events to a form in low-code UI, yet most (75.9\%) of their
questions remain without an accepted answer. We manually label 900  questions
from the posts to determine the prevalence of the topics' challenges across LCSD phases. We find that most of the questions are related to the development phase, and low-code developers also face challenges with automated testing. 
Our study findings offer implications for low-code practitioners,
platform providers, educators, and researchers.
\end{abstract}

\begin{IEEEkeywords}
Low-Code, Issue, Challenge, Empirical Study.
\end{IEEEkeywords}}

%


\maketitle

\IEEEdisplaynontitleabstractindextext

%
\IEEEpeerreviewmaketitle

\section{Introduction}
LCSD is a new paradigm that enables the development of
software applications with minimal hand-coding using visual programming with
graphical interface and model-driven design. LCSD embodies End User Software Programming~\cite{Pane-MoreNatureEUSE-Springer2006} by 
democratizing application development to software practitioners from diverse backgrounds~\cite{di2020democratizing}.  
By facilitating automatic code
generation, the low-code development tools allow developing production-ready applications with minimal coding. It
addresses the gap between domain requirement and developers' understanding that
is a common cause of delayed development in many applications with complex
business logic. The benefits of using  LCSD platforms
also include flexibility and agility, fast development time allowing
quick response to market demands, reduced bug-fixing, lower deployment effort,
and easier maintenance. Hence, the industry of low-code development is gaining
popularity at a rapid pace. According to Forrester
report~\cite{rymer2019forrester}, the  LCSD platform market is expected to be \$21 Billon by
2022. According to Gartner report, by 2024, around 65\% of
large enterprises will use  LCSD platforms to some extent~\cite{wong2019low}.

To date, there are more than 200  LCSD platforms, offered by almost all major companies like Google~\cite{googleappmaker} and Salesforce~\cite{salesforce}. 
Naturally,  LCSD has some unique challenges~\cite{sahay2020supporting}. Wrong
choice of  LCSD application/platforms may cause a waste of
time and resources. There is also concern about the security/scalability of
 LCSD applications~\cite{lowcodetesting}. With interests in  LCSD growing, we observe discussions about  LCSD platforms are becoming prevalent in online developer forums like Stack Overflow (SO). SO is a large online technical Q\&A site with
around 120 million posts and 12 million registered users~\cite{website:stackoverflow}. Several research has been conducted to
analyze SO posts (e.g., big
data~\cite{bagherzadeh2019going}, concurrency~\cite{ahmed2018concurrency}, blockchain~\cite{wan2019discussed}, microservices~\cite{bandeira2019we}). However, we are aware of no
research that analyzed  LCSD discussions on SO, 
although such insight can complement existing  LCSD literature -- which so far has mainly used surveys or controlled studies to understand the needs of low-code practitioners~\cite{lowcodeapp,kourouklidis2020towards,alonso2020towards,lowcodetesting}.  

In this paper, we report an empirical study to understand the types of challenges and topics in  LCSD developer discussions in SO by analyzing all 4.6K SO posts related to the top nine  LCSD platforms at the time of our analysis (according to Gartner). We answer 
three research questions:

\nd\bf{RQ1. What types of topics are discussed about  LCSD in SO?} Given  LCSD is a new paradigm, 
it is necessary to learn about the types of topics  LCSD practitioners discuss in a technical Q\&A site like SO. Therefore, we apply topic modeling algorithm LDA~\cite{blei2003latent} on our dataset of 4.6K posts. We find a total of 13  LCSD topics which are grouped into four categories: Customization of  LCSD UI and Middleware,  
 LCSD Platform Adoption,  LCSD Database Usage, and Third-Party Integration. A majority of the (40\%) questions are asked about the diverse challenges developers face while attempting to customize the user interface (UI) or a service/form provided by an  LCSD platform. This is due to the fact that 
 LCSD platform features are inherently heavy towards a graphical user interface (GUI) in a drag and drop environment. As such, any customization 
of such features that are not directly supported by the  LCSD platforms becomes challenging.     
    
\nd\bf{RQ2. How are the topics distributed across the  LCSD life cycle phases?} Our findings from RQ1 show the unique nature of challenges  LCSD developers face, like customization issues. 
Given the considerable attention towards  LCSD support by software vendors/platforms, the success of the platforms/SDKs can benefit from their effective adoption into the various stages of a software development life cycle (SDLC). For example, if testing of  LCSD 
application cannot be done properly, it is difficult to develop a reliable large-scale  LCSD application. We, therefore, 
need to understand whether and how LCSD developers are discussing the adoption of tools and techniques in  LCSD topic across different SDLC phases. We randomly sampled 900 questions from our dataset and manually analyzed the types of  LCSD challenges developers discussed in the questions. For each question, we label the SDLC phase for which the developer noted the challenge. We found that more than 85\% of the questions revolved around development issues, and it is more or less consistent across all the four topic categories. We also find that testing can be challenging for  LCSD applications due to the graphical nature of the SDKs, which can be hard to debug.   

\nd\bf{RQ3. What  LCSD topics are the most difficult to answer?} Our findings from the above two research questions show that  LCSD developers face challenges more unique to  LCSD platforms (e.g., Customization topics) as well as similar to other domains (e.g., Database topics). Therefore, 
it can be useful to learn what topics are more difficult to get the right answer to and whether the popularity of the topics can suffer due to the observed difficulty. We compute a suite of popularity and difficulty metrics for each topic, like the view count and the percentage of questions without an accepted answer. We find that questions related to the topic ``Dynamic Event Handling'' from the Customization topic category are the most difficult (to get an accepted answer) but also the most popular.

\bf{To the best of our knowledge, ours is the first empirical study of  LCSD and platforms on developer discussions}. 
The findings would help
the research community with a better focus on the specific  LCSD areas. The practitioners can be prepared for difficult areas. Relevant
organizations will be able to design more effective and usable tools for LCSD, increasing their usability.  All stakeholders can work together for improved documentation support. The  LCSD vendors can support increased customization of the  LCSD middleware and UI to make the provided features more usable. 

\nd\bf{Replication Package}: The code and data are shared in \url{https://github.com/disa-lab/LowCodeEmpiricalMSR2021}

\section{Background} \label{sec:background}
\nd\bf{What is an  Low-code Application?} To cater to the demand of the competitive market, business organizations often need to quickly develop and deliver customer-facing applications.  LCSD platform allows the quick translation of the business requirement into a usable software application. It also enables citizen developers of varying levels of software development experience to develop applications using visual tools to design the user interface in a drag-and-drop manner and deploy them easily~\cite{lowcodewiki}.  LCSD is inspired by the model-driven software principle where abstract representations of the knowledge and activities drive the development, rather than focusing on algorithmic computation~\cite{sahay2020supporting}.  LCSD platforms aim to abstract away the complexity of testing, deployment, and maintenance that we observe in traditional software development. Some of the most popular low-code platforms are Appian~\cite{appian}, Google App Maker~\cite{googleappmaker}, Microsoft Powerapps~\cite{powerapps}, and Salesforce Lightning~\cite{salesforce}. 

\begin{figure}[t]
\centering
\includegraphics[scale=.50]{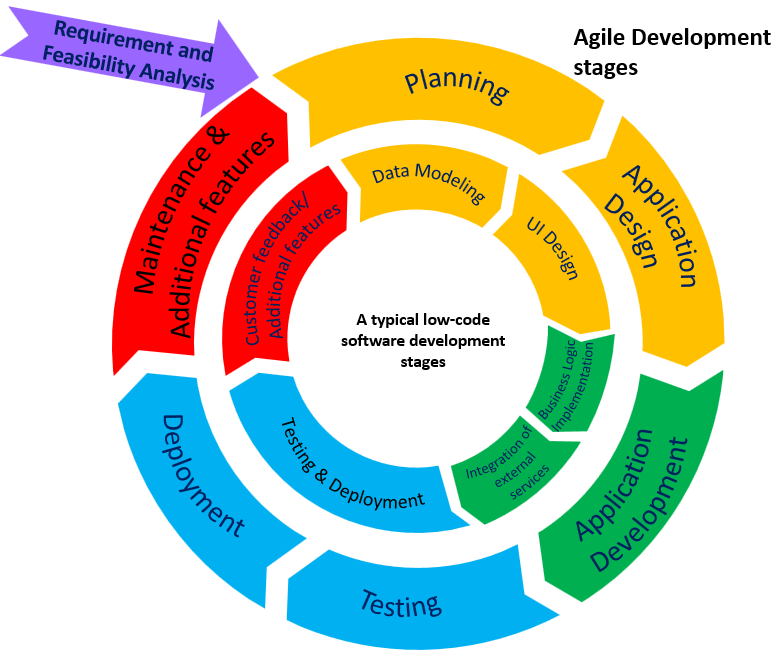}
\caption{Agile methodologies in traditional vs  LCSD development}
\label{fig:low-code-agile}
\vspace{-5mm}
\end{figure}

\nd\bf{Development Phases of an  LCSD Application.} A typical  LCSD application can be built in two ways~\cite{sahay2020supporting}: \begin{inparaenum}
\item ``UI to Data Design'', where developers create UI and then connect the UI to necessary data sources, or \item ``Data to UI'' where the design of the data model is followed by the design of the user interfaces. \end{inparaenum} In both approaches, application logic is implemented, and then third party services and APIs are integrated. APIs are interfaces to reusable software libraries~\cite{Robillard-APIProperty-IEEETSE2012}.  
A major motivation behind  LCSD is to build applications, get reviews from the users, and incorporate those changes quickly~\cite{waszkowski2019low-automating}. As such, the agile development methodology~\cite{beck2001manifesto} and  LCSD can go hand in hand because the fundamental principle and objective are customer satisfaction and continuous incremental delivery. The inner circle of Figure~\ref{fig:low-code-agile} shows the important development phases of an  LCSD application, as outlined in~\cite{sahay2020supporting}. The outer circle of \fig\ref{fig:low-code-agile} shows the phases in a traditional agile software development environment. As  LCSD platforms take care of many of the application development challenges, some of the agile application development phases have shorter time/execution spans in  LCSD compared to traditional software development. 





%

\section{Study Data Collection and Topic Modeling} \label{sec:methodology}
In this Section, we discuss our data collection process to find  LCSD related posts (Section \ref{data_collection}). We then discuss the details of the topic modeling (Section \ref{topic_modeling}).
 
\subsection{Data Collection} \label{data_collection}
We collect  LCSD related SO posts in three steps: \begin{inparaenum}[(1)]
\item Download SO data dump,
\item Identify  LCSD related tag list, and
\item Extract  LCSD related posts from the data dump based on our selected tag list.
\end{inparaenum} We describe the steps below.

\nd\textbf{Step 1: Download SO data dump.} We downloaded SO data dump~\cite{SOdump} of June 2020. We used the contents of Post.xml file, which contained information about each post like the post's unique ID, type (Question or Answer), title, body, associated tags, creation date, view-count, etc. Our data dump included posts from July 2008 to May 2020 and contained around 58,544,636 posts. Out of them, 33.4\% are questions, 66.6\% are answers, and 17.4\% questions had accepted answers.

\nd\textbf{Step 2: Identify low-code tags.}
We need to identify the tags that are related to LCSD in order to extract low-code related posts from SO discussions.
To find relevant tags, we followed a similar procedure used in prior work~\cite{abdellatif2020challenges, ahmed2018concurrency, wan2019discussed, linares2013exploratory}. At Step 1, we identify the initial low-code related tags and call them $T_{init}$. At Step 2, we finalize our low-code tag list following related work~\cite{bagherzadeh2019going, yang2016security}. Our final tag list $T_{final}$ contains 19 tags from the top nine LCDPs. We discuss each step in details below.

(1) Identifying Initial low-code tags.
The SO posts do not have tags like ``low-code'' or ``lowcode''. Instead, we find that  low-code developers use a  LCSD platform name as a tag, e.g., ``appmaker'' for Google Appmaker~\cite{googleappmaker}. Hence, to find relevant tags, first we compile a list of top  LCSD platforms by analysing a list of platforms that are considered as the market leaders in Gartner~\cite{vincent2019magic}, Forrester~\cite{rymer2019forrester}, related research work~\cite{sahay2020supporting}, and other online resources like PC magazine~\cite{pcmag}. We find nine  LCSD platforms are consistently mentioned in the above resources: Zoho Creator~\cite{zohocreator}, Google App Maker~\cite{googleappmaker}, Salesforce Lightning~\cite{salesforce}, Quickbase~\cite{quickbase}, Outsystems~\cite{quickbase}, Mendix~\cite{mendix}, Vinyl~\cite{vinyl}, Appian~\cite{appian}, and Microsoft Powerapps~\cite{powerapps}. 
We thus focus on the discussions of the above nine LCSD platforms in SO. We find one tag per  LCSD platform as the name of the platform (e.g., ``salesforce-lightning''). 
However, upon close inspection of the tags in SO, we found that developers used more than one tag for some of the nine  LCSD platforms. For example, ``Microsoft Powerapps'' has multiple tags (e.g., ``powerapps'', ``powerapps-formula'', ``powerapps-canvas''). At the end of both quantitative analysis and manual validation by four authors, we found a total of 16 tags for the nine  LCSD platforms. We refer to these 16 tags as $T_{init}$. 


(2) Finalizing low-code related tags.
Intuitively, there might be more variations to tags of nine  LCSD platforms other than those in $T_{init}$. We use heuristics from previous related works~\cite{bagherzadeh2019going, yang2016security} to find other relevant tags. First, we denote our entire SO dump data as $Q_{all}$. Second, we extract all the questions $Q$ that contain any tag from $T_{init}$. Third, we create a candidate tag list $T_{candidate}$ using all the tags found in questions $Q$. Fourth, we select significantly relevant tags from  $T_{candidate}$ for our  LCSD discussions. Following related works~\cite{bagherzadeh2019going, yang2016security}, we compute significance and relevance for each tag $t$ in $T_{candidate}$ with respect to $Q$ (our extracted questions that has $T_{init}$ tag) and $Q_{all}$ (i.e., our data dump) as follows,
{ \[
( Significance) \ \ S_{tag} \ =\ \ \frac{\#\ of\ ques.\ with\ the\ tag\ t\ in\ Q}{\ \ \#\ of\ ques.\ with\ the\ tag\ t\ in\ Q_{all}}
\]

\[
( Relevance) \ \ R_{tag} \ =\ \ \frac{\#\ of\ questions\ with\ tag\ t\ in\ Q}{\ \ \#\ of\ questions\ in\ Q}
\]} A tag t is significantly relevant to  LCSD if the $S_{tag}$ and  $R_{tag}$ are higher than a threshold value. We experimented with a wide range of values of $S_{tag}$ and  $R_{tag}$. We found relevant tag set for  LCSD for $S_{tag}$ = 0.2 and $R_{tag}$ = 0.005 . These values are consistent with related work~\cite{bagherzadeh2019going,ahmed2018concurrency}. The final tag list $T_{final}$ contains 19 significantly relevant tags. 

\nd\textbf{Step 3: Extracting low-code related posts.}
An SO question can have at most five tags, and we consider a question as low-code related question if at least one of its tag is in our chosen tag list $T_{final}$. Based on our $T_{final}$ tag set, we found a total of 7,302 posts from our data dump. There were 51.3\% Questions (i.e., 3,747) and 48.7\% Answers (i.e., 3,555) and among them 16.9\% Questions (i.e., 1,236) had accepted answers. SO has a score-based system (upvote and downvote) to ensure the questions are in proper language with necessary information (code samples and error messages), not repeated or off-topic. Here is an example for a question with score ``-4'' where a practitioner is making an API related query in Powerapps(\dq{61147923})\footnote{$Q_i$ and $A_i$ denote a question Q or answer A in SO with an ID $i$} platform. However, it is not clear what the practitioner is asking as the question is poorly written and without any clear example. In order to ensure good quality discussions, we excluded questions that had a negative score which resulted in 6,982 posts containing 51.5\% Questions (i.e., 3,597 ) and 48.5\% Answers (i.e., 3,385). Following previous research~\cite{bagherzadeh2019going, rosen2016mobile, barua2014developers}, we excluded unaccepted answers and only considered accepted answers for our dataset. Hence, our final dataset $B$ contained  4,785 posts containing 3,597 non-negative scored questions and 1,188 accepted answers. 

\subsection{Topic Modeling} \label{topic_modeling}
We produce  LCSD topics from our extracted posts in three steps: \begin{inparaenum}[(1)]
\item Preprocess the posts, 
\item Find optimal number of topics, and
\item Generate topics.
\end{inparaenum} We discuss the steps below.

\nd\textbf{Step 1. Preprocess the posts.} For each post text, we remove noise following related works~\cite{abdellatif2020challenges,bagherzadeh2019going,barua2014developers}. First, we remove the code snippets from the body, which is inside \textless code\textgreater \textless /code\textgreater\ tag, HTML tags such as (\textless p\textgreater \textless /p\textgreater, \textless a\textgreater \textless /a\textgreater, \textless li\textgreater \textless /li\textgreater\ etc), and URLs. Then we remove the stop words such as ``the'', ``is'', ``are'', punctuation marks, numbers, non-alphabetical characters using the stop word list from MALLET~\cite{mccallum2002mallet}, NLTK~\cite{loper2002nltk}, and our custom low-code specific (i.e.,  LCSD platform names) stop word list. After this, we use porter stemmer~\cite{ramasubramanian2013effective} to get the stemmed representations of the words e.g., ``wait'', ``waits'', ``waiting'', and ``waited'' - all of which are stemmed to base form ``wait''.

\nd\textbf{Step 2. Finding the optimal number of topics.}  After the prepossessing, we use Latent Dirichlet Allocation~\cite{blei2003latent} and the MALLET tool~\cite{mccallum2002mallet} to find out the  LCSD related topics in SO discussions. We follow similar studies in Software engineering research using topic modeling~\cite{arun2010finding, asuncion2010software, yang2016security, bagherzadeh2019going,abdellatif2020challenges}. Our goal is to find the optimal number of topics $K$ for our dataset $B$ so that the \textit{coherence} score, i.e., encapsulation of underlying topics, is high. We use Gensim package~\cite{rehurek2010software} to determine the coherence score following previous works~\cite{uddin2017automatic, roder2015exploring}. We experiment with different values of $K$ that range from \{5, 6, 7, .., 29, 30\} and for each value, we run  MALLET LDA on our dataset for 1000 iterations~\cite{bagherzadeh2019going}. Then we observe how the coherence score is changing with respect to $K$. We pick the topic model with the highest coherence score. Choosing the right value of $K$ is important because, for smaller values of $K$, multiple real-world concepts merge, and for a large value of $K$, a topic breaks down. For example, in our result, the highest coherence score 0.50  for $K$ = 7 and $K$ = 13. We choose $K$ = 13 as it captures our underlying topics better. For $K$ = 7, we find that it merges the topics ``Dynamic Event Handling'', ``Dynamic Content Display'', and ``Dynamic Form Controller''. MALLET also uses two hyper-parameters, $\alpha$ and $\beta$, to distribute words and posts across the topics. Following the previous works~\cite{bagherzadeh2019going, ahmed2018concurrency, bajaj2014mining, rosen2016mobile}, we use the standard values $50/K$ and 0.01 for hyper-parameters $\alpha$ and $\beta$ in our experiment.  

\nd\textbf{Step 3. Generating topics.} Topic modeling is a method of extracting a set of topics by analysing a collection of documents without any predefined taxonomy. Each document has a probability distribution of topics, and every topic has a probability distribution of a set of related words~\cite{barua2014developers}. We produced 13 topics using the above LDA configuration on our dataset $B$. Each topic model offers a list of top $N$ words and a list of $M$ posts associated with the topic. In our settings, a topic consists of 30 most frequently co-related words, which represents a concept. Each post had a correlation score between 0 to 1, and following the previous work~\cite{wan2019discussed}, we assign a document with a topic that it correlates most.

\section{Empirical Study} \label{sec:results}
We answer three research questions by analyzing LCSD discussions in SO.
The RQ1 aims to understand the types of topics discussed in
SO about LCSD (\sec\ref{sec:rq-topic-type}). The RQ2 aims to understand how each topic is discussed across different stages of low-code SDLC (software development life cycle) (\sec\ref{sec:rq-topic-challenges}). The RQ3 offers insight into the challenges of LCSD 
developers across the observed LCSD topics based on the difficulty of getting answers (\sec\ref{sec:rq-topic-difficulty}).

\subsection{What types of topics are discussed about LCSD? (RQ1)}\label{sec:rq-topic-type}


\subsubsection{Approach}
We get 13 low-code related topics from our LDA topic modeling, as discussed in \sec\ref{sec:methodology}. We use card sorting~\cite{fincher2005making} to label these topics  following previous
works~\cite{bagherzadeh2019going, ahmed2018concurrency, yang2016security, rosen2016mobile, abdellatif2020challenges}.
In open card sorting, there is no predefined list of labels. To label a topic, 
we used the top 30 words for the topic and a random sample of at least 20 questions that are assigned to the topic. Four of the authors participated in the labelling process.
Each author assigns a label for each topic and discusses with each other until
there is an agreement. The authors reached an agreement after around 15
iterations of meetings over Skype and email and labeled the 13 topics from the
LDA output discussed in Section \ref{sec:background}. After the labeling of the topics, we grouped them into higher categories. For
example, UI Adaptation and Dynamic Form Controller are related to UI design, and
so we group them into a group named UI. In the same way, Dynamic Event Handling,
Dynamic Content Display, and Dynamic Content Binding topics are related to the
middleware feature of low-code development platforms, and so we put these three
topics into Middleware sub-category. We repeat this process until we can not
find any more higher-level group. For example, the above mentioned two
categories UI and Middleware belong to the application customization task where
the developers customize the UI or the business logic of the application
according to their need. Hence, we put them under a high-level category named
Customization. Similarly, we put Access Control \& Security into Configuration sub-category. Then we put Configuration sub-category and Client Server Comm \& IO topic under Platform Adaptation high-level category.


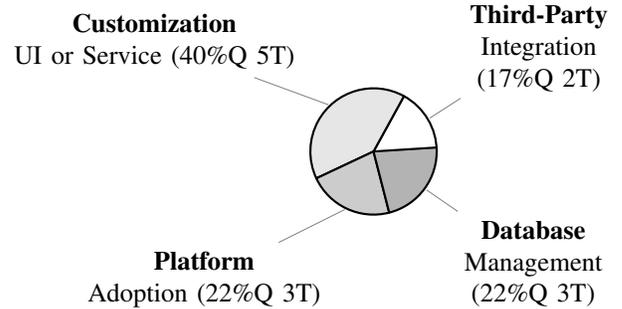
\begin{figure}[t]
	\centering\begin{tikzpicture}[scale=0.28]-
    \pie[
        /tikz/every pin/.style={align=center},
        text=pin, number in legend,
        explode=0.0,
        color={black!0, black!10, black!20, black!30,, black!40},
        ]
        {
            17/\bf{Third-Party}\\Integration\\ (17\%Q 2T),
            40/\bf{Customization}\\UI or Service (40\%Q 5T),
            22/\bf{Platform}\\ Adoption (22\%Q 3T),
            22/\bf{Database}\\Management\\ (22\%Q 3T)
        }
    \end{tikzpicture}
	\caption{Distribution of questions (Q) and topics (T) per topic category}
	\vspace{-5mm}
	\label{fig:distribution_of_questions_topic_pie_chart}
\end{figure}

\begin{figure}[t]
\centering
\includegraphics[scale=0.32]{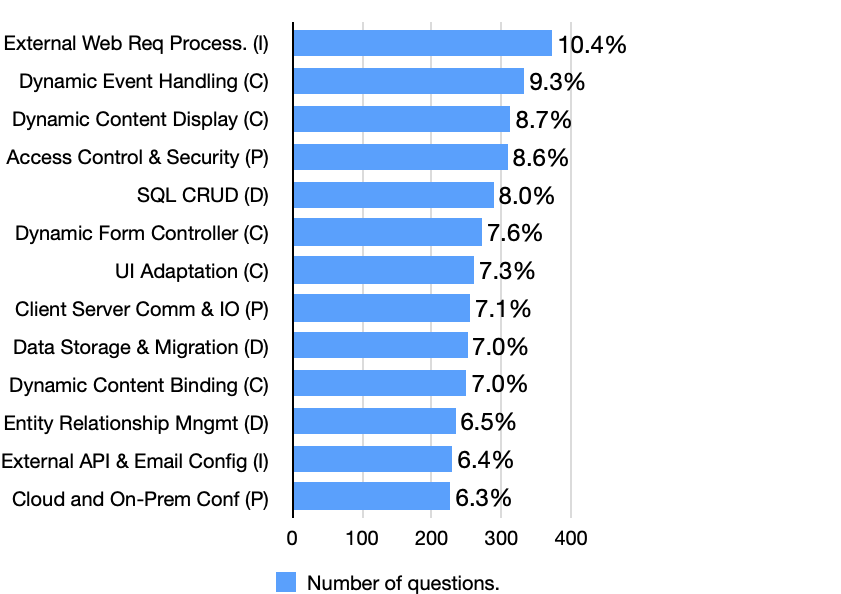}
\caption{Distribution of questions by low-code Topics (C = Customization Category, I = Integration, P = Platform Adoption, D = Database)}
\label{fig:distribution_of_questions_topic_bar_chart}
\end{figure}
\subsubsection{Results}

\begin{figure}[t]
\centering
\includegraphics[scale=0.48]{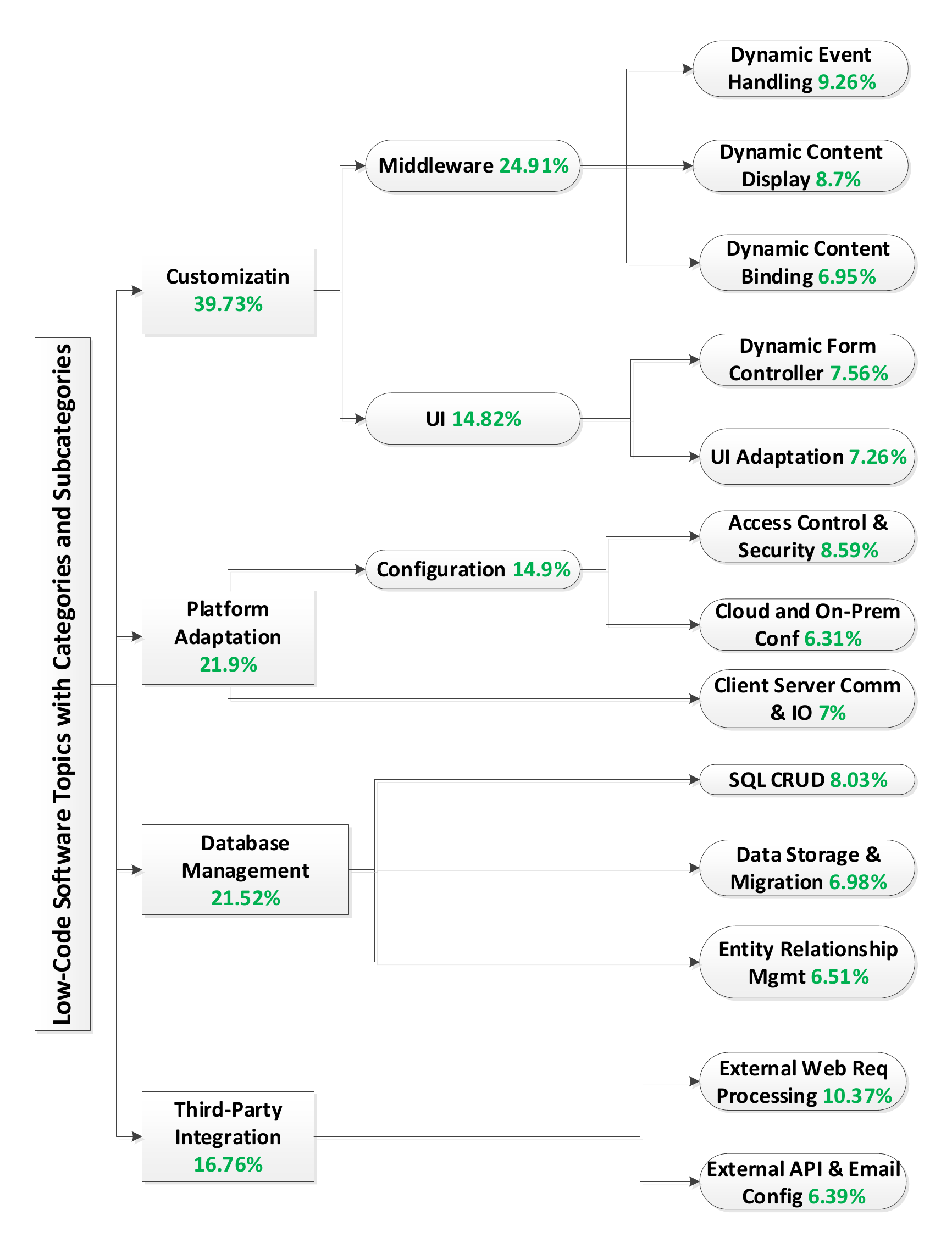}
\caption{ Low-code topics categories and sub-categories}
\label{fig:taxonomy_TM}
\end{figure}

We find 13 LCSD topics that we group into four high-level categories: \textbf{Customization, Platform Adoption, Database Management, and Third-Party Integration.} \fig\ref{fig:distribution_of_questions_topic_pie_chart} shows the distribution of
questions and topics among the four high-level categories. The \textit{Customization} category 
covers the highest percentage of questions
and number of topics (40\% questions and five topics), followed by \textit{ Platform Adoption} (22\%
questions and three topics), \textit{Database Management} (33\% questions and three topics), and \textit{Integration}
(17\% questions and two topics). \fig\ref{fig:distribution_of_questions_topic_bar_chart} sorts the 13 topics based on number of questions. The
\textit{``External Web Req Processing topic''} covers the most questions regarding queries and discussions related
to the integration of third party services/APIs (10.4\%). The \textit{``Dynamic Event Handling''} has the second most questions (9.3\%) related to the implementation of business logic.

In Figure \ref{fig:taxonomy_TM}, we group the 13 topics into four high-level categories. The categories are sorted according to the number of
questions belonging to them. For example, the topmost category \textit{Customization} has the most number of questions under them. Each category may consist of some
sub-categories. For example, \textit{Customization} category contains two sub-categories:
\begin{inparaenum}[(1)]
\item \textit{Middleware}, and
\item \textit{UI}
\end{inparaenum}. Each sub-category contains one or more topics. For example,
Middleware sub-category consists of three topics: \begin{inparaenum}[(1)] \item
Dynamic Event Handling, \item Dynamic Content Display, and \item Dynamic Content Binding\end{inparaenum}. Each sub-category and topic are organized according to
their distribution of questions. For example, under Customization (40\%)
category, \textit{Middleware} (25\%) is followed by \textit{UI} (15\%) sub-category. In the same
way, ``Dynamic Event Handling'' (9.3\%) topic is followed by ``Dynamic Content
Display'' (8.7\%) topic based on their question distribution.



\nd\bf{$\bullet$ Customization} is the largest category, with 40\% of the SO questions and five topics. It contains discussions about business
logic implementation, input and form validation, linking the UI to the backend
storage via dynamic content binding, a drop-down menu with predefined value,
formatting date and time, drop-down widgets, etc. It contains two sub-categories
of topics: 
\begin{inparaenum}[(1)] 
    \item \textit{Middleware} sub-category covers discussions on the middlewares that
provide support for system integration, the connection between UI and storage layer, etc., and 
    \item \textit{UI} sub-category contains discussions
on drag-and-drop UI and form design and also customization of UI components.
\end{inparaenum}

\begin{inparaenum}[(i)]
    \item \bf{{Middleware (25\% questions)}} sub-category contains three topics:
    \begin{inparaenum}[(1)] \item \it{Dynamic Event Handling (9.26\%)} has
    discussions about handling user interaction events, accessing input
    value after form submission (\dq{43096166}), and rendering chart in the
    canvas(\dq{56154215}).
    \item \it{Dynamic Content Display (8.70\%)} is about dynamically displaying
    items on the page (\dq{53648077}), displaying content based on previous action, and creating and accessing gallery from multiple data sources (\dq{51764889}).
    \item \it{Dynamic Content Binding (6.95\%)} is about updating views when some
    other values get changed (\dq{59932262}) and building process based on some values
    on the form (\dq{61282976}).
    \end{inparaenum} 
    
    \item \bf{{UI (15\% questions)}} sub-category contains two topics:
    \begin{inparaenum}[(1)] \item \it{Dynamic Form Controller (7.56\%)} contains
    discussions related to the design of forms with predefined values and the implementation of multi-select and customized drop-down values (\dq{44013975}), adding event-listeners to text
    widget (\dq{46038130}), etc.
    \item \it{UI Adaptation (7.26\%)} is about designing and customizing the user
    interface, resizing screen, etc.(\dq{34515865}).
    \end{inparaenum} 
\end{inparaenum}



\nd\bf{$\bullet$ Platform  Adoption} is the second-largest category, with 22\% of the questions. It contains discussions about generic query on LCSD platform features
and support, role management, SDLC management tools (e.g., scrum, agile), cloud setup and configuration, deployment issues, etc. The category has three topics. The topic, \it{Client Server Comm \& IO (7.09\%)}, contains discussions on client-server architecture (\dq{54900592}), debugging server-side
scripts (\dq{55283256}), and general debugging queries on error messages or
unexpected output (\dq{50936643}). The other two topics are grouped under Configuration sub-category. 

\begin{inparaenum}[(i)]
    \item \bf{{Configuration (15\% questions)}} contains discussions on LCSD platform configuration on access control and cloud-based setup. There are two topics:
    \begin{inparaenum}[(1)] \item \it{Access Control \& Security (8.59\%)} is about
    discussion on role-based access control to tasks (\dq{51431318}), configuration of existing
    authentication mechanism (Azure Active Directory configuration) (\dq{61734680}).
    \item \it{Cloud and On-Prem Conf (6.31\%)} contains discussion on the proper configuration parameters and 
    guidelines to connect to the cloud/on-prem databases (\dq{55207558} and \dq{45740520}).
    \end{inparaenum} 
\end{inparaenum}



\nd\bf{$\bullet$ Database} is tied to the second biggest category, with 22\% of the questions. It contains discussions about database connection, SQL CRUD
operations, import/export existing data, etc. There are three topics:
\begin{inparaenum}[(1)] 
    \item \it{SQL CRUD (8.03\%)} contains discussions on SQL
        query (\dq{49051500}, \dq{59852901}) and table joining (\dq{37707699}), 
    \item \it{Data Storage \& Migration (6.98\%)} discusses the upload
        and storing of files on the server (\dq{49666940}), moving files from one platform
        to another (\dq{55726281}), conversion of large CSV files to excel sheet
        (\dq{50977178}), etc., 
    \item \it{Entity Relationship Management (6.51\%)} contains discussion on
        relational database design  (\dq{51881224}) and platform support/limitations
        on relational database (\dq{58935331}).
\end{inparaenum} 



\nd\bf{$\bullet$ Integration} is the smallest category, with 17\% of the questions and two topics. It contains discussions about email server configuration, integration
of external services, OAuth, fetching and parsing data, etc. It contains two
topics: \begin{inparaenum}[(1)] \item \it{External Web Req Processing (10.37\%)}
contains posts regarding API integration, parsing and debugging the responses of
REST APIs (\dq{21314917}), and OAuth (\dq{56873258}), some general query on
networking protocols such as HTTP, REST API (\dq{48628269}), etc.,
\item \it{External API \& Email Config (6.39\%)} is about configuration, sending
or forwarding emails (\dq{34085695}), configuration error (\dq{31501424}), how
to use a generic programming language to send an email (\dq{36341976}), creating
managing calendar events (\dq{46738962}), etc.
\end{inparaenum}

\begin{tcolorbox}[flushleft upper,boxrule=1pt,arc=0pt,left=0pt,right=0pt,top=0pt,bottom=0pt,colback=white,after=\ignorespacesafterend\par\noindent]
\noindent\textbf{Summary of RQ1.} We found 13 topics in our SO dataset relevant to low-code
software development.  The topics belong to four categories: Customization,
Platform Adoption, Database, and Integration. Customization category has the
most number of questions, followed by Platform Adoption, Database, and
Integration. Out of the topics, ``External Web Request Processing'' topic under
the Integration category constitutes the highest number of questions (10.4\%),
followed by the topic ``Dynamic Event Handling'' (9.3\%) under the Customization
category.
\end{tcolorbox}

\subsection{How are the topics distributed across the  LCSD life cycle phases? (RQ2)}\label{sec:rq-topic-challenges}



\subsubsection{Approach}
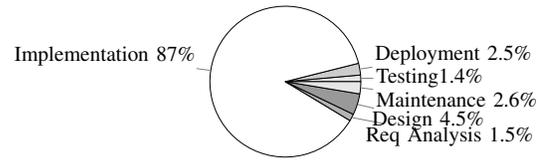
\begin{figure}[t]
	\centering
	\resizebox{2.8in}{!}{
	\begin{tikzpicture}
    \pie[
        explode=0.0, text=pin, number in legend, sum = auto, 
        color={black!10, black!20, black!0, black!30,black!40},
        ]
        {
            1.4/\Huge{Testing}1.4\%,
            2.5/\Huge{Deployment} 2.5\%,
            87/\Huge{Implementation} 87\%,
            1.5/\Huge{Req Analysis} 1.5\%,
            4.5/\Huge{Design} 4.5\%,
            2.6/\Huge{Maintenance} 2.6\%
        }
    \end{tikzpicture}
    }
	\caption{Distribution of questions (Q) per SDLC phase}
	\label{fig:distribution_of_SDLC_pie_chart}
\end{figure}

Agile software development methodology~\cite{beck2001manifesto} has six SDLC phases:  \begin{inparaenum}[(1)]
\item Requirement Analysis \& Planning, 
\item Application Design,
\item Implementation, 
\item Testing,
\item Deployment, and
\item Maintenance\end{inparaenum}.
Our final dataset $B$ contained 3,597 questions. We needed at least 347 questions to produce a statistically significant sample, with a 95\% confidence level and 5 confidence interval. As noted in Fig. \ref{fig:distribution_of_SDLC_pie_chart}, some phases have a low number of questions. Thus we wanted to find more examples of those phrases in our sample. As such, we analyzed 900 randomly sampled questions. The labelling of each question to determine the precise SDLC phases was conducted by several co-authors in joint discussion sessions spanning over 80 person-hours. Eventually, we manually labelled 916 questions, out of which 16 were considered invalid because they did not have any specific LCSD discussions. For example, a new practitioner is tasked with finding the right LCSD platform during the planning stage of his/her LCSD application. The practitioner queries, ``Are there any serious pitfalls to Outsystems Agile Platform?'' (\dq{3016015}). We thus assign the SDLC phase for it as ``Requirement Analysis \& Planning''. Another question asks, ``Google  App  Maker  app  not  working  after  deploy'' (\dq{42506938}). We label the SDLC phase as ``Deployment''.

\begin{table}[t]
  \centering
   \caption{Distribution (frequency) of LCSD topics per SDLC phase. Each colored bar denotes a phase 
   (Black = Requirement Analysis \& Planning, Green = Application Design, Magenta = Implementation, Red = Testing, Blue = Deployment, Orange = Maintenance)}
    \resizebox{\columnwidth}{!}{%
    \begin{tabular}{lr}
    \toprule{}
    \textbf{Topics} & \textbf{Development Phases Noted in \#Questions}\\
    \midrule
    \textbf{Customization (351, 39\%)} &  \\
    Dynamic Event Handling	& \sixbars{0}{0}{73}{7}{1}{0}		\\
    Dynamic Content Display & \sixbars{0}{2}{78}{1}{0}{0}	\\
    Dynamic Form Controller	& \sixbars{0}{2}{67}{0}{0}{0}	\\
    UI Adaptation	& \sixbars{0}{1}{59}{1}{1}{2} \\
    Dynamic Content Binding	& \sixbars{0}{5}{49}{0}{0}{2} \\
    \midrule
    \textbf{Platform Adoption (189, 21\%)} &  \\
    Access Control \& Security	& \sixbars{4}{3}{42}{2}{14}{12} \\
    Client Server Comm \& IO	& \sixbars{0}{0}{54}{2}{1}{0}  \\
    Cloud and On-Prem Conf & \sixbars{8}{12}{28}{1}{4}{2} \\
    \midrule 
    \textbf{Database (214, 23.7\%)} &  \\
    SQL CRUD	& \sixbars{1}{4}{74}{0}{0}{0} \\
    Data Storage \& Migration	& \sixbars{0}{3}{63}{0}{0}{2} \\
    Entity Relationship Mngmt	& \sixbars{1}{5}{58}{0}{0}{3} \\
    \midrule
    \textbf{Integration (145, 16.1\%)} &  \\
    Ext Web Req Processing & \sixbars{0}{3}{99}{0}{0}{0}	\\
    Ext API \& Email Config & \sixbars{0}{0}{41}{0}{1}{1}\\
    \bottomrule
    \end{tabular}%
   }
  \label{tab:topicSDLC}%
\end{table}%

\subsubsection{Results} Figure \ref{fig:distribution_of_SDLC_pie_chart} shows that the Implementation phase is found in 87\% of the 900 questions we studied, followed by Design (4.5\%), Maintenance (2.6\%) and Deployment (2.4\%) phases. This is not surprising, given that SO is a technical Q\&A site and developers use the forum to find solutions to their technical problems. Table \ref{tab:topicSDLC} shows the distribution of our 13 LCSD topics over six low-code SDLC phase based on our analysis of 900 SO questions. Besides the dominance of development phase related questions across all topics, we find the presence of other phases (e.g., testing, deployment) in customization and platform adoption topics.

\bf{Requirement Analysis \& Planning (14, 1.5\%).} 
Requirement analysis is the process of developing software according to the expectation of the users. 
During planning, the feasibility, timeline, dependability, potential complexity/risks are analyzed and planned by paying attention to the operational aspects. The LCSD platforms usually provide requirement management tools that allow developers to collect data, customize checklists, import user stories into sprint plans. At this phase, the developers usually face enquiries about cost, learning curve, LCSD platform's support for faster application development, deployment, and maintenance features to choose the right platform for their business need. For example, in this popular question, a new practitioner is asking for some drawbacks on some potential pitfalls for a particular LCSD platform, e.g., "Are there any serious pitfalls to Outsystems Agile Platform?" (\dq{3016015}). A developer from that platform provider suggests using the platform to build an application and decide for himself as it is hard to define what someone might consider a pitfall. 





\bf{Application Design (40, 4.5\%).} In this phase, the design specification is made based on the requirements of the application. All the key stakeholders review and approve this considering application architecture, modularity, and extensibility. The LCSD developers face challenges regarding data storage design, drag and drop UI design, using on-prem data-sources with the LCSD platform (e.g., ``Can AppMaker be used with SQL Server'' (\dq{55220499})), migrating existing data to LCSD platform (\dq{46421271}), designing a responsive web page (e.g., ``Incorporating responsive design in App Maker'' (\dq{52744026})).


\bf{Implementation (785, 87.3\%).} At this phase, actual application development begins. LCSD developers face a wide range of challenges when trying to customize the UI, implement business logic, integrate third-party modules, debug and test the implemented functionalities, read the incomplete or incorrect documentation, etc. Developers ask application customization and UI customization questions like ``How do I change timezone in AppMaker Environment?'' (\dq{47731051}), drop-down menu customization (e.g., ``powerapps: populate drop down list from another datasource'' (\dq{40159662})). There are many queries (17\%) regarding external services or API integration. Many of these questions are relevant to a task-based tutorial on how to use a REST API and process its response (e.g., ``How to upload files and attachments to the subject record using REST API?'' (\dq{61143493})). The root cause of many of these issues is incomplete or incorrect documentation. For example, in \dq{46241015}, a practitioner is querying about the integration of Zoho CRM and says that s/he does not understand the sample code. In \dq{34510911}, a practitioner asks for a sample code to convert a web page to a PDF. Clearly, the documentation is not sufficient for a smooth transition for entry-level practitioners.




\bf{Testing (14, 1.5\%).} 
The testing process of LCSD varies from traditional software. It usually takes less testing in LCSD platforms because the platform management team test and monitors the modules provided. Unit testing carries less importance than tradition development because the components are already unit tested, and developers usually integrate those using drag-and-drop fashion. Many platforms provide custom unit testing features for the application logic code added by the developers. The platform providers recommend running a security audit to check if there is a potential data exposure. Most of the issues faced in this phase are related to browser compatibility, lack of proper documentation to run automated tests, test coverage, issues using the third party functional testing tools like Selenium (\dq{61210424}), etc. Practitioners make general queries about running tests on low-code platforms (\dq{46669690}) and errors while running test (\dq{47254010}).




\bf{Deployment (22, 2.5\%).} LCSD platforms aim to make the \textit{deployment and maintenance} phase smooth. Many platforms provide Application Life-Cycle Management tools to develop, debug, deploy, and maintain the staging and production server. However, LCSD developers still face challenges regarding deployment configuration issues (\dq{46369742}), version control, DNS configuration, performance issues, accessibility issues (i.e., sharing public URL of the application (\dq{44136328}, \dq{53884162})), DNS configuration, etc. For example, in this discussion, a developer was having trouble accessing the app after deployment (e.g., ``Google App Maker app not working after deploy'' (\dq{42506938})). In the accepted answer, a community member provides a detailed description to achieve that and points out the lack of \textit{Official Documentation} for such a crucial task. There are a few queries about deploying application with custom URL, i.e., the domain name (e.g., ``How to make friendly custom URL for deployed app'' (\dq{47194231})). In this case, it was difficult because the platform did not have native support. 





\bf{Maintenance (24, 2.6\%).} At this stage, the application is released and needs maintenance support. The users sometimes find bugs that were not caught before and sometimes want new features that may spawn a new software development life cycle in agile or incremental development methodology. In this phase, developers face challenges regarding bugs in a low-code platform and difficulties in using different application maintenance features provided by the platform, such as event monitoring, collaboration, application design reusability, etc. Different LCSD platforms provide features on developers' role management, dashboard, and event monitoring. For example, in this question, a practitioner queries about the feasibility of role-based access control in an LCSD platform: ``PowerApps : Implementing Role Based Security In Your PowerApps App'' (\dq{52762374}). LCSD developers also find it difficult to determine/update versions of LCSD platforms (\dq{45209796}). 

\begin{tcolorbox}[flushleft upper,boxrule=1pt,arc=0pt,left=0pt,right=0pt,top=0pt,bottom=0pt,colback=white,after=\ignorespacesafterend\par\noindent]
\noindent\textbf{Summary of RQ2.} We 
randomly sampled 900 questions from our dataset and manually examined the SDLC phase of the questions. We found an overwhelming majority of (85\%) implementation phase-related questions in SO. Non-coding questions are generally discouraged in SO, so we found very few questions related to other phases (e.g., requirement analysis, deployment, etc.). We also find that LCSD developers find testing to be challenging for LCSD application due to the graphical nature of the SDKs, which can be hard to debug.
\end{tcolorbox}

\subsection{What LCSD topics are the most difficult to answer? (RQ3)}\label{sec:rq-topic-difficulty}

\begin{table*}[t]
  \centering
   \caption{Low-code software development topics, their popularity, and difficulty}
    \begin{tabular}{llrrrrrrr}
    \toprule{}
    \multirow{2}{*}{\textbf{Topic}} & \multirow{2}{*}{\textbf{Category}} & \multicolumn{4}{c}{\textbf{Popularity score}} & \multicolumn{3}{c}{\textbf{Difficulty score}} \\
    \cmidrule{3-9}
    {} & {}  & \textbf{Avg view} & \textbf{Avg fav.} & \textbf{Avg score} & \textbf{Avg \#ans} & \textbf{W/O any ans.} & \textbf{W/O acc. ans.} & \textbf{Med. Hrs acc.} \\
    \midrule

   Dynamic Event Handling	&	Customization &	833 &	1.23 &	0.54 & 0.86	& 37.2\% & 75.9\% &	9.8 \\
   Ext. Web Req Processing & Integration & 785 &	1.24 &	0.62 & 1.04	& 23.3\% &	68.1\% &	15.7 \\ 
   Ext. API \& Email Config &		Integration &	764 &	1.23 &	0.83 &	0.91 &	34.3\% & 70.4\% &	12.7 \\
   Dynamic Content Display &	Customization 	& 722 &	0.86 & 0.48 & 0.99 &	17.6\%	& 65.2\%	& 14.8 \\
   Cloud and On-Prem Conf	&	Adoption &	578 &	1.18 &	0.85 & 1.01 &	23.8\% &	66.5\% &	16.8 \\  
   Dynamic Form Controller	&	Customization  &	566 &	0.85 &	0.45 & 1.07	& 15.1\% &	57.4\% &	4.2 \\
    UI Adaptation	&	Customization  &	536 &	0.95 & 	0.46 & 0.88 &	30.3\%	& 68.2\% &	6.1 \\ 
    Dynamic Content Binding	&	Customization &	507 &	1.16 &	0.36 & 0.94	& 22.4\% &	67.2\% &	24.9 \\
   Entity Relationship Mngmt	&	Database &	485  &	1.09 &	0.48 &	0.93 &	29.5\% & 62.4\% &	6.9 \\  
    Data Storage \& Migration	&	Database &	472 &	1.07 &	0.60 & 0.89 &	25.1\% &	69.3\% &	14.8 \\
    Client Server Comm \& IO	&	Adoption &	408	& 1.11 &	0.54 & 0.86	& 31.8\% &	67.8\% &	12.7 \\ 
    SQL CRUD	&	Database	& 359 &	1.04 &	0.45 & 0.92 &	22.1\%  &	60.2\% &	7.6 \\
   Access Control \& Security	&	Adoption &	301 &	0.97 &	0.49 & 0.93 &	26.2\% & 69.9\% &	12.6 \\
    \midrule
    \multicolumn{2}{l}{\textbf{Average}}  & \textbf{572} & \textbf{1.1} & \textbf{0.5} & \textbf{0.9} & 26\% &\textbf{67\%} & 12.3 \\
    \bottomrule
    \end{tabular}%
  \label{tab:topicPopularityDifficulty}%
\end{table*}%

\begin{table}[t]
  \centering
  \caption{Correlation between the topic popularity and difficulty}
    \begin{tabular}{lrrr}\toprule
    coefficient/p-value & \bf{View} & \bf{Favorites} & \bf{Score}\\ \midrule
    \bf{\% without acc. ans.} &   0.154/0.51    & 0.348/0.10      &  0.379/0.08 \\
    \bf{Hrs to acc. ans.} &    0.077/0.77   &   0.400/0.06     & 0.431/0.04  \\
    \bf{\% without any answer} &   0.077/0.77    & 0.374/0.08      &  0.275/0.20 \\
    \bottomrule
    \end{tabular}%
  \label{tab:correlation}%
\end{table}



\subsubsection{Approach} For each topic, we compute the difficulty of getting answers under a topic using three metrics: \begin{inparaenum}[(1)] 
\item Percentage of questions without any answer at all,
\item Percentage of questions without an accepted answer,
\item Average median time needed to get an accepted answer.
\end{inparaenum} In the same way we use the following four popularity metrics: \begin{inparaenum}[(1)]
\item Average number of views, \item Average number of favourites, \item Average
score, and \item Average number of answers\end{inparaenum}. Then we aim to determine the correlation between the difficulty and
popularity of the topics. The first two difficulty metrics and the first three popularity metrics are used in several previous studies~\cite{bagherzadeh2019going, abdellatif2020challenges, ahmed2018concurrency}. We use the Kendall Tau correlation measure~\cite{Kendall-TauMetric-Biometrica1938} to
find the correlation between topic popularity and topic difficulty. SO does not
provide the data across a time-series for all metrics such as view count, score, etc. As such, our analysis offers as of time insight. 

\subsubsection{Results}
Table \ref{tab:topicPopularityDifficulty} shows the topic difficulty using the three metrics, as noted above. 
These metrics allow us to understand the difficulty of getting a working solution~\cite{bagherzadeh2019going, abdellatif2020challenges}.
The \textit{Dynamic Event Handling} under Customization has the highest average view count and the highest percentage of questions (76\%) without an accepted answer. \textit{Dynamic Content Binding} from the same category also has the highest median hours to get an accepted answer. Many questions in this topic relate to business logic customization in an LCSD platform, which is not familiar to other developers. For example, \dq{51443599} asks, ``How to add more data an array of objects in a lightning component?''. This question has been asked around two years ago, viewed around 11K times and still active. It has three answers, but none of them is marked as accepted. Whereas, \textit{Dynamic Form Controller} under Customization is the least difficult topic concerning the percentage of the questions without an accepted answer (57\%) and median time (4.2 hours) to get a solution. It is because questions related to form design and validation have good \textit{community support}. The same is also true for \textit{SQL CRUD}, which has around 7.6 median hours for accepted answers.

The \textit{External Web Req Processing} under
 \textit{Integration} has the highest percentage of questions and the highest average
favourite count. The \textit{Cloud and On-Prem Conf}
under  \textit{Platform} adoption have the highest average score. The
discussions are about setting and maintaining cloud configuration and migrating
on-prem data to the server in this topic. For example, in \dq{44727285}, a practitioner is
asks ``Migrate Salesforce data from one org to another'', and in \dq{3016015}, a
new practitioner is making a general low-code platform related query ``Are
there any serious pitfalls to Outsystems Agile Platform?''. This question has
a very high score because lots of new low-code platform developers have faced this
issue. \textit{Access Control \& Security} under Platform Adoption is the least
popular topic in terms of average view count. In this topic, the questions are about deploying the application, its security and access control. For example, in
\dq{17886545}, a practitioner says, ``Can't Connect to SalesForce in C\#''. He also
explains there is a security token error, and in the documentation, it is not
mentioned how to configure that. Many of these questions are not general, and
so it has a low average view-count.

\nd\textbf{Correlation between topic difficulty and popularity.} 
In Table \ref{tab:correlation}, we present nine
correlation measures using three difficulty and popularity metrics from Table
\ref{tab:topicPopularityDifficulty}. Our result shows no
statistically significant correlation between the popularity and difficulty
metrics since for eight out of nine correlation values are \textgreater~0.05. Therefore, we cannot say that the least popular topics are the most difficult
ones and vice versa. For example, Access Control \& Security is one of
the least popular topics but considered among the most difficult topics (Table
\ref{tab:topicPopularityDifficulty}). However, this observation does not hold for
Dynamic Event Handling, which is the most popular and among
the most difficult topics.  


\begin{tcolorbox}[flushleft upper,boxrule=1pt,arc=0pt,left=0pt,right=0pt,top=0pt,bottom=0pt,colback=white,after=\ignorespacesafterend\par\noindent]
\noindent\textbf{Summary of RQ3.} We compute seven popularity and difficulty metrics for each topic using metrics such as  
view count, percentage of questions without an accepted answer, etc. We find that questions related to the topic ``Dynamic Event Handling'' from the 
customization category are the most difficult (to get an accepted answer) but also the most popular (average view count). Overall, we do not see any significant correlation (positive/negative) between topic difficulty and popularity metrics. 
\end{tcolorbox}

\section{Discussions} \label{sec:discussion}
In this section, we discuss the evolution of low-code and LCSD related discussions with respect to different topics. Then we discuss the implications of our findings.

\subsection{Evolution of LCSD topics}

\begin{figure}[t]
\centering
\includegraphics[scale=0.38]{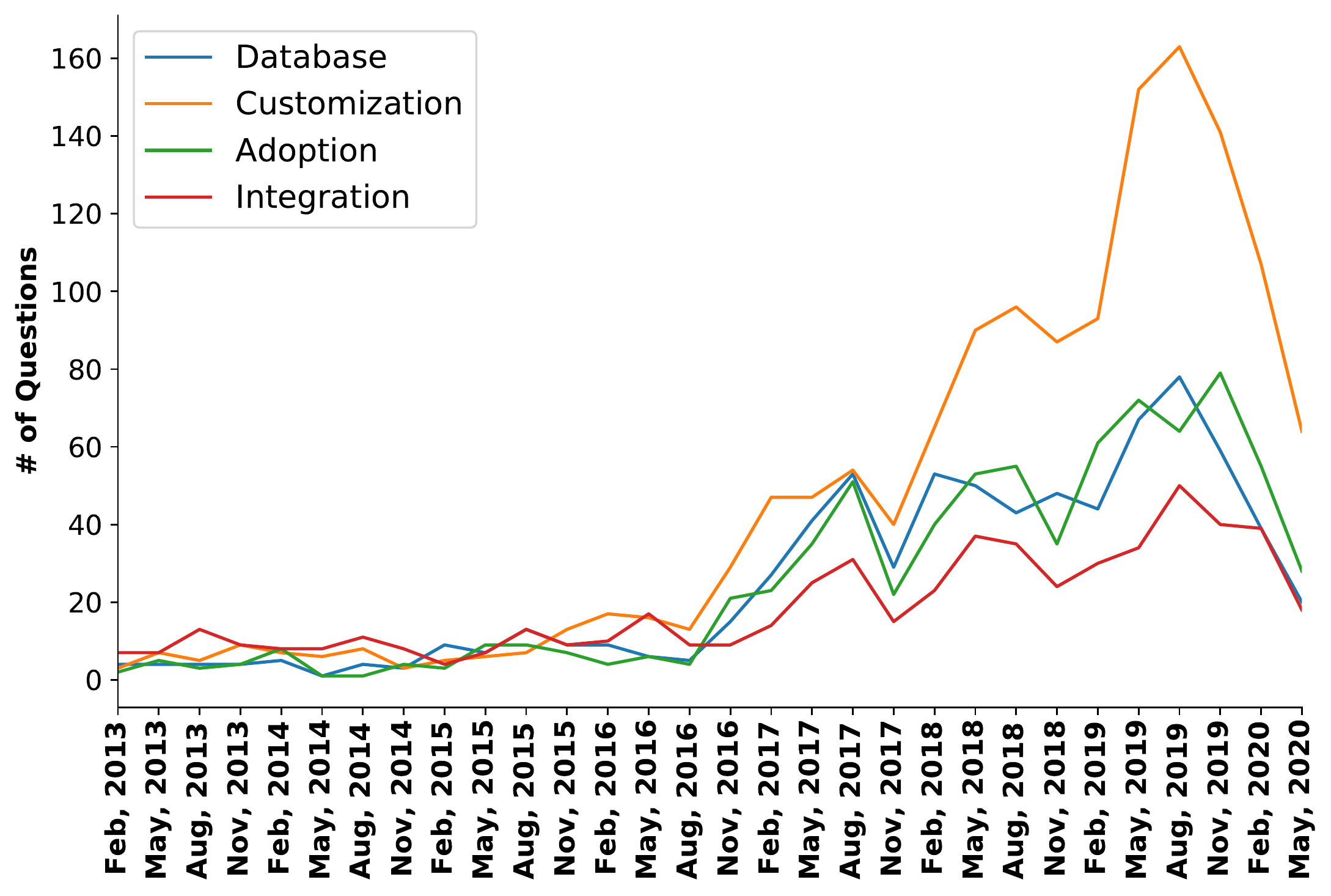}
\caption{Low-code topic category evolution over time.}
\label{fig:trend_questions_per_topic_category}
\end{figure}

We measure the growth of our four high-level topic categories over time to better understand the evolution of LCSD. We measure the absolute growth, i.e., the total number of questions in a category over time. Figure ~\ref{fig:trend_questions_per_topic_category} shows that all four of our topic categories are increasing monotonically.
This trend indicates that the LCSD approach is gaining more community attention over time, especially after 2016. 

\begin{figure}[t]
\centering
\includegraphics[scale=0.38]{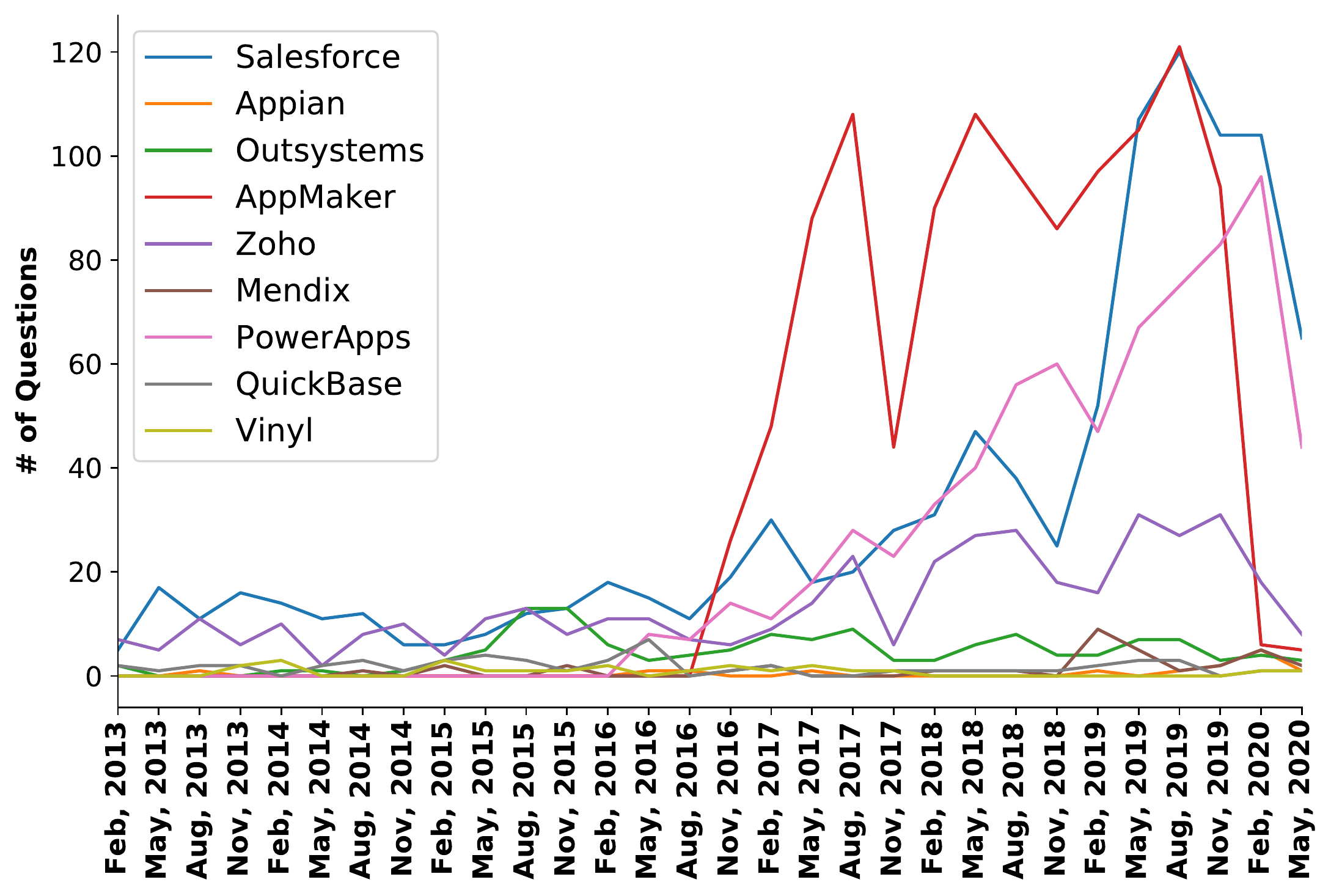}
\caption{Low-code Platforms evolution over time}
\label{fig:trend_questions_per_platform}
\end{figure}

We further analyze two sudden swings in the number of questions. First, we find an increase of questions in every category after mid-2016, especially for questions about the \textit{Customization} category. Google released App Maker~\cite{googleappmaker} for public use in 2016, which introduced many discussions on LCSD customization. \fig\ref{fig:trend_questions_per_platform}  confirms it and shows a spike in questions about Google App Maker during that time. The second case is that at the beginning of 2020, there is a sharp decline in SO discussions. In Jan 2020, Google announced that they would no longer release new features for Google App Maker and discontinue it by 2021~\cite{google-disc}. It created unrest among the developer community as they were trying to verify this information (\dq{59947680}) and to explore alternatives (e.g., \dq{59985750}). \fig\ref{fig:trend_questions_per_platform} also shows the sudden drop in the number of questions asked about Google App Maker starting Jan 2020.


\subsection{Implications of Findings} 


\begin{figure}[t]
\vspace{-5mm}
\includegraphics[scale=0.36]{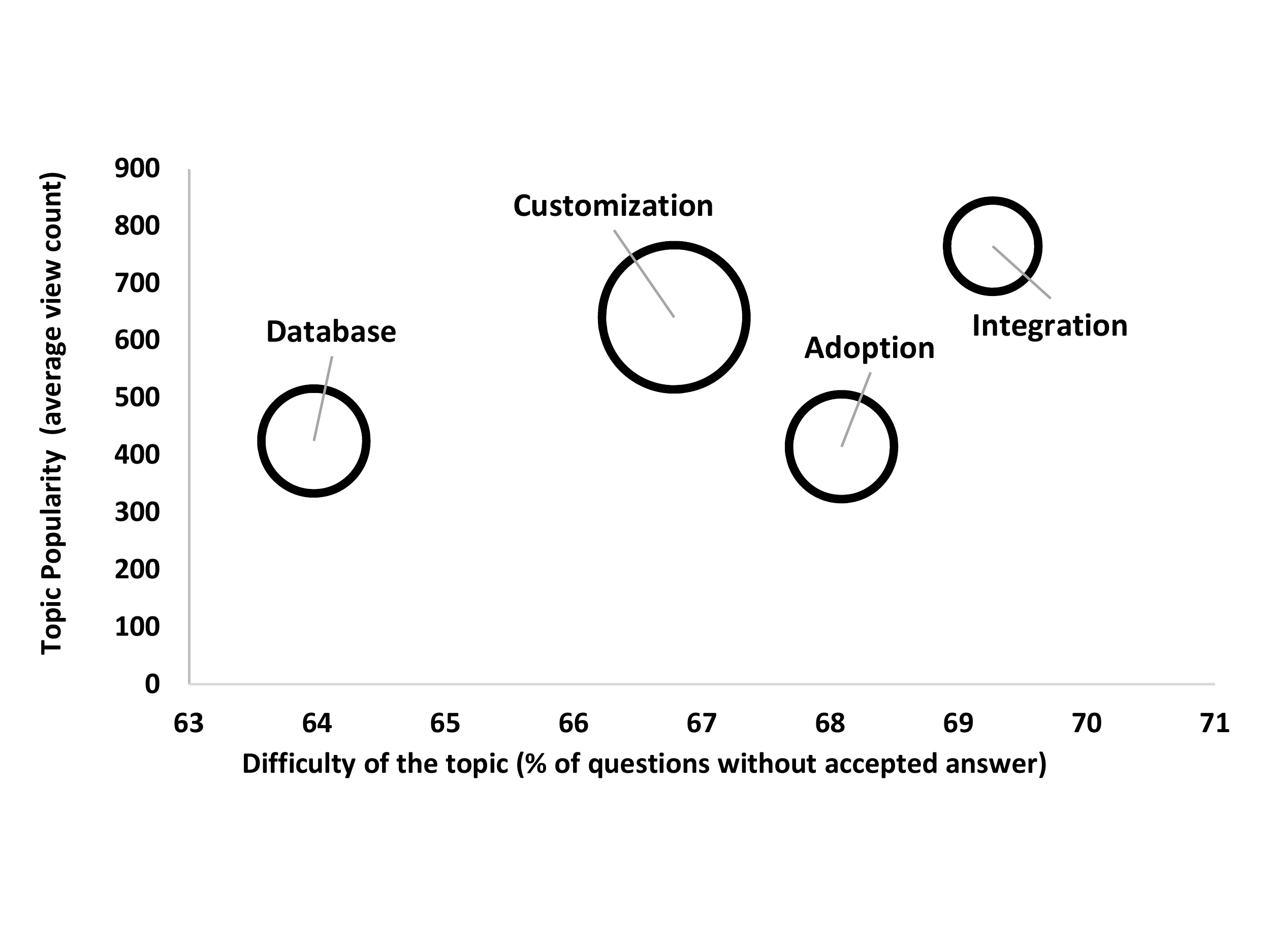}
\vspace{-15mm}
\caption{Low-code topic categories popularity vs. difficulty}
\label{fig:bubble_diff_pop_per_category}
\end{figure}


This study can help the low-code community to focus on the pressing issues on the
LCSD paradigm. We discuss the implications of our study findings by stakeholders below.

\bf{\ul{ LCSD Platform Providers.}} 
In order to better understand the issues of LCSD, we present a bubble chart \ref{fig:bubble_diff_pop_per_category} that presents the positions
of low-code categories in terms of popularity vs. difficulty. In this study, we
use the average number of view-count and percentage of questions without accepted
answers as a proxy for the topic category popularity and difficulty,
respectively~\cite{abdellatif2020challenges}. The size of the bubble depends on
the number of questions for that particular topic category.
Figure \ref{fig:bubble_diff_pop_per_category} shows that \textit{Integration} is
the most popular as well as most difficult topic category. On the other hand,
\textit{Database} remains the least difficult category due to the superior database support by  LCSD platforms. As shown in Figure \ref{fig:bubble_diff_pop_per_category},
\textit{Customization} is the largest and prevalent low-code topic category.
Many new practitioners make queries regarding LCSD platforms,
learning resources, basic application and UI customization, and how to get
started with this new emerging technology. We find that \textit{Documentation}
related queries are both very popular and difficult. Our
findings also suggest that many practitioners still face challenges during testing and debugging. Consequently, many of
the questions on this topic remain unanswered. It reveals that to ensure smooth adoption of the LCSD platform, the platform providers should provide better and effective documentation and provide learning resources to reduce entry-level barriers and smooth out the learning curve. 

\textbf{\ul{ LCSD Practitioners/Developers.}} LCSD abstractions and the platform's
feature limitations sometimes make it very difficult to customize and debug. Our
finding shows that the practitioners find third party service \textit{Integration}
and \textit{Platform Feature} category most difficult. It provides valuable
insights for project managers to manage resources better (i.e., human resources and
development time). 
LCSD platform enables practitioners with diverse experience to contribute to the development process even without a software development
background. However, our finding shows that practitioners find debugging, application
accessibility, and documentation challenging. Hence, the practitioners should take
the necessary steps to understand the tradeoffs LCSD platforms' features deeply. The project manager should adopt specific strategies to learn to customize, debug, and test the application.

\textbf{\ul{ LCSD Researchers \& Educators.}} We find that
the LCSD paradigm's challenges can be different from traditional software development~\cite{sahay2020supporting}. Researchers can focus on the most popular
and difficult topic category \textit{Integration} and
develop a set of metrics to automatically detect documentation quality on
third-party service APIs. Simultaneously, researchers can study how to
provide better tools for practitioners to customize the application.
Security is an open research opportunity for such platforms as a security
vulnerability in such platforms or frameworks could compromise millions of applications and users~\cite{lin2020software}. Researchers can come
up with better testing approaches to ensure faster development and dependability. Educators can also benefit from the results presented in Figure
\ref{fig:bubble_diff_pop_per_category} to prioritize their focus on different topics such as \textit{Database, Customization, and Third-party API Integration}.

\section{Threats to Validity}\label{subsec:validity}

\bf{Internal validity} threats relate to the authors'
bias while conducting the analysis. We mitigate the bias in
our manual labeling of topics and  LCSD phases by consulting the labels among multiple authors. Four of the authors
actively participated in the labelling process. The third author reviewed the final labels and refined the labels by consulting with the first author. \bf{Construct Validity} threats 
relate to the errors that may occur in data collection like, identifying relevant LCSD tags. To mitigate this, we examine all the tags that we find in the low-code related questions. Then we expanded our tag list using state-of-art approach~\cite{bagherzadeh2019going,abdellatif2020challenges,ahmed2018concurrency, rosen2016mobile}. Another potential threat is the topic modeling technique, where we choose $K$ = 13 as the optimal number of topics for our dataset $B$. This optimal number of topics have a direct impact on the output of LDA. We experimented with different values of $K$ following related works~\cite{abdellatif2020challenges, bagherzadeh2019going}. We used the coherence score and manual examination to find $K$'s optimal that gives us the most relevant and generalized low-code related topics. \bf{External Validity} threats relate to the generalizability of our findings. Our study is based on data from developers' discussion on SO. However, there are other forums  LCSD developers may use to discuss. Nevertheless, we believe using SO's data provides us with generalizability because SO is a widely used Q\&A platform for developers. To ensure good quality discussion, we only use posts with non-negative scores. However, we also believe this study can be complemented by including discussions from other forums, surveying and interviewing low-code developers.

\section{Related Work} \label{sec:related_work}
\nd\bf{Research on low-code development.} LCSD is a relatively new technology, and there are only a few research works in this domain. There is some research on how this emerging technology can 
be used in different software applications~\cite{lowcodeapp} or for automating business process in manufacturing~\cite{waszkowski2019low-automating}. Sipio et al.~\cite{di2020democratizing} 
present the benefits and future potential of LCSD by sharing their experience of building a custom recommendation system in the LCSD
platform. Kourouklidis et al.~\cite{kourouklidis2020towards} discuss
the low-code solution to monitor the machine learning model's performance. Sahay
et al. survey LCDP and compare different LCDPs based on
their helpful features and functionalities~\cite{sahay2020supporting}. Khorram
et al.~\cite{lowcodetesting} analyse commercial LCSD platforms and present a
list of features and testing challenges. Ihirwe et al.~\cite{lowcodeIot}
analyse 16 LCSD platforms and identifies what IoT application-related features and services each platform provides. All these research works
compare LCSD platforms and their support on the different types
of applications~\cite{alonso2020towards}. To the best of our knowledge, ours is the first empirical study of LCSD and platforms based on developer discussions.

\nd\bf{Topic Modeling in Software Engineering.} Our motivation to use topic modeling to understand  LCSD discussions stems from
existing research in software engineering that shows that topics generated from
textual contents can be a good approximation of the underlying
\it{themes}~\cite{Chen-SurveyTopicInSE-EMSE2016,Sun-SoftwareMaintenanceHistoryTopic-CIS2015,Sun-ExploreTopicModelSurvey-SNPD2016}.
Topic models are used recently to understand software
logging~\cite{Li-StudySoftwareLoggingUsingTopic-EMSE2018} and previously for
diverse other tasks, such as concept and feature
location~\cite{Cleary-ConceptLocationTopic-EMSE2009,Poshyvanyk-FeatureLocationTopic-TSE2007},
traceability linking (e.g.,
bug)~\cite{Rao-TraceabilityBugTopic-MSR2011,asuncion2010software},
to understand software and source code history
evolution~\cite{Hu-EvolutionDynamicTopic-SANER2015,Thomas-SoftwareEvolutionUsingTopic-SCP2014,Thomas-EvolutionSourceCodeHistoryTopic-MSR2011},
to facilitate code search by categorizing
software~\cite{Tian-SoftwareCategorizeTopic-MSR2009}, to refactor software code
base~\cite{Bavota-RefactoringTopic-TSE2014}, as well as to explain software
defect~\cite{Chen-SoftwareDefectTopic-MSR2012}, and various software maintenance
tasks~\cite{Sun-SoftwareMaintenanceTopic-IST2015,Sun-SoftwareMaintenanceHistoryTopic-CIS2015}.
The SO posts are subject to several studies on various aspects
of software development using topic modeling, such as what developers are
discussing in general~\cite{barua2014developers} or about a
particular aspect, e.g., concurrency~\cite{ahmed2018concurrency}, big
data~\cite{bagherzadeh2019going}, chatbot development~\cite{abdellatif2020challenges}.
We are aware of no previous research on understanding
the  LCSD discussions in SO.
\section{Conclusions} \label{sec:conclusion}
LCSD is a new paradigm that enables the development of
software applications with minimal hand-coding using visual programming. We present an empirical study that provides insights into the types of topics low-code developers discuss in Stack Overflow (SO). We find 13 low-code topics in our dataset of 4.6K SO posts (question + accepted answers). The posts are collected based on 19 SO tags belonging to the popular nine  LCSD platforms during our analysis. We categorize them into four high-level groups, namely Customization, Platform Adoption, Database, and Integration. Our findings reveal that developers find the external API Integration topic category the most challenging and the Database category least difficult. Dynamic Event Handling is the most popular, as well as the most challenging topic. We find a severe lack of good tutorial based documentation that deters the smooth adaptation of LCSD. We hope that all of these findings will help various  LCSD stakeholders (e.g.,  LCSD platforms, practitioners, SE researchers) to take necessary actions to address the various  LCSD challenges. Since the growth indicates that this technology is likely to be widely adopted by various companies for their internal and customer-facing applications, platform providers should address the prevailing developers' challenges.



\balance
\bibliographystyle{plainnat}

{\small 
\bibliography{consolidated_new}

\begin{thebibliography}{63}
\providecommand{\natexlab}[1]{#1}
\providecommand{\url}[1]{\texttt{#1}}
\expandafter\ifx\csname urlstyle\endcsname\relax
  \providecommand{\doi}[1]{doi: #1}\else
  \providecommand{\doi}{doi: \begingroup \urlstyle{rm}\Url}\fi

\bibitem[Abdellatif et~al.(2020)Abdellatif, Costa, Badran, Abdalkareem, and
  Shihab]{abdellatif2020challenges}
Ahmad Abdellatif, Diego Costa, Khaled Badran, Rabe Abdalkareem, and Emad
  Shihab.
\newblock Challenges in chatbot development: A study of stack overflow posts.
\newblock In \emph{Proceedings of the 17th International Conference on Mining
  Software Repositories}, MSR '20, page 174–185, New York, NY, USA, 2020.
  Association for Computing Machinery.
\newblock ISBN 9781450375177.
\newblock \doi{10.1145/3379597.3387472}.
\newblock URL \url{https://doi.org/10.1145/3379597.3387472}.

\bibitem[Ahmed and Bagherzadeh(2018)]{ahmed2018concurrency}
Syed Ahmed and Mehdi Bagherzadeh.
\newblock What do concurrency developers ask about? a large-scale study using
  stack overflow.
\newblock In \emph{Proceedings of the 12th ACM/IEEE International Symposium on
  Empirical Software Engineering and Measurement}, ESEM '18, New York, NY, USA,
  2018. Association for Computing Machinery.
\newblock ISBN 9781450358231.
\newblock \doi{10.1145/3239235.3239524}.
\newblock URL \url{https://doi.org/10.1145/3239235.3239524}.

\bibitem[Alonso et~al.(2020)Alonso, Abreu, Nunes, Vieira, Santos, Soares, and
  Pereira]{alonso2020towards}
Ana~Nunes Alonso, Jo{\~a}o Abreu, David Nunes, Andr{\'e} Vieira, Luiz Santos,
  T{\'e}rcio Soares, and Jos{\'e} Pereira.
\newblock Towards a polyglot data access layer for a low-code application
  development platform.
\newblock \emph{arXiv preprint arXiv:2004.13495}, 2020.

\bibitem[appian()]{appian}
appian.
\newblock {Appian platform overview}.
\newblock {Available: \url{https://www.appian.com/}}.
\newblock [Online; accessed 5-January-2021].

\bibitem[Arun et~al.(2010)Arun, Suresh, Madhavan, and Murthy]{arun2010finding}
Rajkumar Arun, Venkatasubramaniyan Suresh, CE~Veni Madhavan, and MN~Narasimha
  Murthy.
\newblock On finding the natural number of topics with latent dirichlet
  allocation: Some observations.
\newblock In \emph{Pacific-Asia conference on knowledge discovery and data
  mining}, pages 391--402. Springer, 2010.

\bibitem[Asuncion et~al.(2010)Asuncion, Asuncion, and
  Taylor]{asuncion2010software}
Hazeline~U Asuncion, Arthur~U Asuncion, and Richard~N Taylor.
\newblock Software traceability with topic modeling.
\newblock In \emph{2010 ACM/IEEE 32nd International Conference on Software
  Engineering}, volume~1, pages 95--104. IEEE, 2010.

\bibitem[Bagherzadeh and Khatchadourian(2019)]{bagherzadeh2019going}
Mehdi Bagherzadeh and Raffi Khatchadourian.
\newblock Going big: A large-scale study on what big data developers ask.
\newblock In \emph{Proceedings of the 2019 27th ACM Joint Meeting on European
  Software Engineering Conference and Symposium on the Foundations of Software
  Engineering}, ESEC/FSE 2019, pages 432--442, New York, NY, USA, 2019. ACM.

\bibitem[Bajaj et~al.(2014)Bajaj, Pattabiraman, and Mesbah]{bajaj2014mining}
Kartik Bajaj, Karthik Pattabiraman, and Ali Mesbah.
\newblock Mining questions asked by web developers.
\newblock In \emph{Proceedings of the 11th Working Conference on Mining
  Software Repositories}, pages 112--121, 2014.

\bibitem[Bandeira et~al.(2019)Bandeira, Medeiros, Paixao, and
  Maia]{bandeira2019we}
Alan Bandeira, Carlos~Alberto Medeiros, Matheus Paixao, and Paulo~Henrique
  Maia.
\newblock We need to talk about microservices: an analysis from the discussions
  on stackoverflow.
\newblock In \emph{2019 IEEE/ACM 16th International Conference on Mining
  Software Repositories (MSR)}, pages 255--259. IEEE, 2019.

\bibitem[Barua et~al.(2014)Barua, Thomas, and Hassan]{barua2014developers}
Anton Barua, Stephen~W Thomas, and Ahmed~E Hassan.
\newblock What are developers talking about? an analysis of topics and trends
  in stack overflow.
\newblock \emph{Empirical Software Engineering}, 19\penalty0 (3):\penalty0
  619--654, 2014.

\bibitem[Bavota et~al.(2014)Bavota, Oliveto, Gethers, Poshyvanyk, and
  Lucia]{Bavota-RefactoringTopic-TSE2014}
Gabriele Bavota, Rocco Oliveto, Malcom Gethers, Denys Poshyvanyk, and Andrea~De
  Lucia.
\newblock Methodbook: Recommending move method refactorings via relational
  topic models.
\newblock \emph{IEEE Transactions on Software Engineering}, 40\penalty0
  (7):\penalty0 671--694, 2014.

\bibitem[Beck et~al.(2001)Beck, Beedle, Van~Bennekum, Cockburn, Cunningham,
  Fowler, Grenning, Highsmith, Hunt, Jeffries, et~al.]{beck2001manifesto}
Kent Beck, Mike Beedle, Arie Van~Bennekum, Alistair Cockburn, Ward Cunningham,
  Martin Fowler, James Grenning, Jim Highsmith, Andrew Hunt, Ron Jeffries,
  et~al.
\newblock Manifesto for agile software development.
\newblock 2001.

\bibitem[Blei et~al.(2003)Blei, Ng, and Jordan]{blei2003latent}
David~M. Blei, Andrew~Y. Ng, and Michael~I. Jordan.
\newblock Latent dirichlet allocation.
\newblock \emph{Journal of Machine Learning Research}, 3\penalty0
  (4-5):\penalty0 993--1022, 2003.

\bibitem[Chen et~al.(2012)Chen, Thomas, Nagappan, and
  Hassan]{Chen-SoftwareDefectTopic-MSR2012}
Tse-Hsun Chen, Stephen~W. Thomas, Meiyappan Nagappan, and Ahmed~E. Hassan.
\newblock Explaining software defects using topic models.
\newblock In \emph{9th working conference on mining software repositories},
  pages 189--198, 2012.

\bibitem[Chen et~al.(2016)Chen, Thomas, and
  Hassan]{Chen-SurveyTopicInSE-EMSE2016}
Tse-Hsun~(Peter) Chen, Stephen~W. Thomas, and Ahmed~E Hassan.
\newblock A survey on the use of topic models when mining software
  repositories.
\newblock \emph{Empirical Software Engineering}, 21\penalty0 (5):\penalty0
  1843--1919, 2016.

\bibitem[Cleary et~al.(2009)Cleary, Exton, Buckley, and
  English]{Cleary-ConceptLocationTopic-EMSE2009}
Brendan Cleary, Chris Exton, Jim Buckley, and Michael English.
\newblock An empirical analysis of information retrieval based concept location
  techniques in software comprehension.
\newblock \emph{Empirical Software Engineering}, 14:\penalty0 93--130, 2009.

\bibitem[Di~Sipio et~al.(2020)Di~Sipio, Di~Ruscio, and
  Nguyen]{di2020democratizing}
Claudio Di~Sipio, Davide Di~Ruscio, and Phuong~T Nguyen.
\newblock Democratizing the development of recommender systems by means of
  low-code platforms.
\newblock In \emph{Proceedings of the 23rd ACM/IEEE International Conference on
  Model Driven Engineering Languages and Systems: Companion Proceedings}, pages
  1--9, 2020.

\bibitem[Exchange(2020)]{SOdump}
Stack Exchange.
\newblock { Stack exchange data dump }.
\newblock {Available: \url{https://archive.org/details/stackexchange}}, 2020.
\newblock [Online; accessed 5-January-2021].

\bibitem[Fincher and Tenenberg(2005)]{fincher2005making}
Sally Fincher and Josh Tenenberg.
\newblock Making sense of card sorting data.
\newblock \emph{Expert Systems}, 22\penalty0 (3):\penalty0 89--93, 2005.

\bibitem[Fryling(2019)]{lowcodeapp}
Meg Fryling.
\newblock Low code app development.
\newblock \emph{J. Comput. Sci. Coll.}, 34\penalty0 (6):\penalty0 119, April
  2019.
\newblock ISSN 1937-4771.

\bibitem[googleappmaker()]{googleappmaker}
googleappmaker.
\newblock {Google App Maker platform overview}.
\newblock {Available: \url{https://developers.google.com/appmaker}}.
\newblock [Online; accessed 5-January-2021].

\bibitem[googledisc()]{google-disc}
googledisc.
\newblock {Google App Maker will be shut down on January 19, 2021}.
\newblock
  \url{https://workspaceupdates.googleblog.com/2020/01/app-maker-update.html}.
\newblock [Online; accessed 5-January-2021].

\bibitem[Hu et~al.(2015)Hu, Sun, Lo, and
  Li]{Hu-EvolutionDynamicTopic-SANER2015}
Jiajun Hu, Xiaobing Sun, David Lo, and Bin Li.
\newblock Modeling the evolution of development topics using dynamic topic
  models.
\newblock In \emph{IEEE 22nd International Conference on Software Analysis,
  Evolution, and Reengineering}, pages 3--12, 2015.

\bibitem[Ihirwe et~al.(2020)Ihirwe, Di~Ruscio, Mazzini, Pierini, and
  Pierantonio]{lowcodeIot}
Felicien Ihirwe, Davide Di~Ruscio, Silvia Mazzini, Pierluigi Pierini, and
  Alfonso Pierantonio.
\newblock Low-code engineering for internet of things: A state of research.
\newblock In \emph{Proceedings of the 23rd ACM/IEEE International Conference on
  Model Driven Engineering Languages and Systems: Companion Proceedings},
  MODELS '20, New York, NY, USA, 2020. Association for Computing Machinery.
\newblock ISBN 9781450381352.
\newblock \doi{10.1145/3417990.3420208}.
\newblock URL \url{https://doi.org/10.1145/3417990.3420208}.

\bibitem[Kendall(1938)]{Kendall-TauMetric-Biometrica1938}
M.~G. Kendall.
\newblock A new measure of rank correlation.
\newblock \emph{Biometrika}, 30\penalty0 (1):\penalty0 81--93, 1938.

\bibitem[Khorram et~al.(2020)Khorram, Mottu, and Suny\'{e}]{lowcodetesting}
Faezeh Khorram, Jean-Marie Mottu, and Gerson Suny\'{e}.
\newblock Challenges \& opportunities in low-code testing.
\newblock In \emph{Proceedings of the 23rd ACM/IEEE International Conference on
  Model Driven Engineering Languages and Systems: Companion Proceedings},
  MODELS '20, New York, NY, USA, 2020. Association for Computing Machinery.
\newblock ISBN 9781450381352.
\newblock \doi{10.1145/3417990.3420204}.
\newblock URL \url{https://doi.org/10.1145/3417990.3420204}.

\bibitem[Kourouklidis et~al.(2020)Kourouklidis, Kolovos, Matragkas, and
  Noppen]{kourouklidis2020towards}
Panagiotis Kourouklidis, Dimitris Kolovos, Nicholas Matragkas, and Joost
  Noppen.
\newblock Towards a low-code solution for monitoring machine learning model
  performance.
\newblock In \emph{Proceedings of the 23rd ACM/IEEE International Conference on
  Model Driven Engineering Languages and Systems: Companion Proceedings}, pages
  1--8, 2020.

\bibitem[Li et~al.(2018)Li, Chen, Shang, and
  Hassan]{Li-StudySoftwareLoggingUsingTopic-EMSE2018}
Heng Li, Tse-Hsun~(Peter) Chen, Weiyi Shang, and Ahmed~E. Hassan.
\newblock Studying software logging using topic models.
\newblock \emph{Empirical Software Engineering}, 23:\penalty0 2655–2694,
  2018.

\bibitem[Lin et~al.(2020)Lin, Wen, Han, Zhang, and Xiang]{lin2020software}
Guanjun Lin, Sheng Wen, Qing-Long Han, Jun Zhang, and Yang Xiang.
\newblock Software vulnerability detection using deep neural networks: a
  survey.
\newblock \emph{Proceedings of the IEEE}, 108\penalty0 (10):\penalty0
  1825--1848, 2020.

\bibitem[Linares-V{\'a}squez et~al.(2013)Linares-V{\'a}squez, Dit, and
  Poshyvanyk]{linares2013exploratory}
Mario Linares-V{\'a}squez, Bogdan Dit, and Denys Poshyvanyk.
\newblock An exploratory analysis of mobile development issues using stack
  overflow.
\newblock In \emph{2013 10th Working Conference on Mining Software Repositories
  (MSR)}, pages 93--96. IEEE, 2013.

\bibitem[Loper and Bird(2002)]{loper2002nltk}
Edward Loper and Steven Bird.
\newblock Nltk: the natural language toolkit.
\newblock \emph{arXiv preprint cs/0205028}, 2002.

\bibitem[lowcodewiki()]{lowcodewiki}
lowcodewiki.
\newblock {Low-code development platform }.
\newblock {Available:
  \url{https://en.wikipedia.org/wiki/Low-code_development_platform}}.
\newblock [Online; accessed 5-January-2021].

\bibitem[McCallum(2002)]{mccallum2002mallet}
Andrew~Kachites McCallum.
\newblock Mallet: A machine learning for language toolkit.
\newblock \emph{http://mallet. cs. umass. edu}, 2002.

\bibitem[mendix()]{mendix}
mendix.
\newblock {Mendix platform overview}.
\newblock {Available: \url{https://www.mendix.com/}}.
\newblock [Online; accessed 5-January-2021].

\bibitem[Overflow(2020)]{website:stackoverflow}
Stack Overflow.
\newblock \emph{Stack Overflow Questions}.
\newblock \url{https://stackoverflow.com/questions/}, 2020.
\newblock Last accessed on 14 November 2020.

\bibitem[Pane and Myers(2006)]{Pane-MoreNatureEUSE-Springer2006}
John Pane and Brad Myers.
\newblock \emph{More Natural Programming Languages and Environments}, pages
  31--50.
\newblock Springer, 10 2006.
\newblock ISBN 978-1-4020-4220-1.
\newblock \doi{10.1007/1-4020-5386-X_3}.

\bibitem[pcmag()]{pcmag}
pcmag.
\newblock {The Best Low-Code Development Platforms}.
\newblock {Available:
  \url{https://www.pcmag.com/picks/the-best-low-code-development-platforms}}.
\newblock [Online; accessed 5-January-2021].

\bibitem[Poshyvanyk et~al.(2007)Poshyvanyk, Guéhéneuc, Marcus, Antoniol, and
  Rajlich]{Poshyvanyk-FeatureLocationTopic-TSE2007}
Denys Poshyvanyk, Yann-Gaël Guéhéneuc, Andrian Marcus, Giuliano Antoniol,
  and Václav~T Rajlich.
\newblock Feature location using probabilistic ranking of methods based on
  execution scenarios and information retrieval.
\newblock \emph{IEEE Transactions on Software Engineering}, 33\penalty0
  (6):\penalty0 420--432, 2007.

\bibitem[powerapps()]{powerapps}
powerapps.
\newblock {Microsoft power apps platform overview}.
\newblock {Available: \url{https://powerapps.microsoft.com/en-us/}}.
\newblock [Online; accessed 5-January-2021].

\bibitem[quickbase()]{quickbase}
quickbase.
\newblock {Quickbase platform overview}.
\newblock {Available:
  \url{https://www.quickbase.com/product/product-overview}}.
\newblock [Online; accessed 5-January-2021].

\bibitem[Ramasubramanian and Ramya(2013)]{ramasubramanian2013effective}
C~Ramasubramanian and R~Ramya.
\newblock Effective pre-processing activities in text mining using improved
  porter’s stemming algorithm.
\newblock \emph{International Journal of Advanced Research in Computer and
  Communication Engineering}, 2\penalty0 (12):\penalty0 4536--4538, 2013.

\bibitem[Rao and Kak(2011)]{Rao-TraceabilityBugTopic-MSR2011}
Shivani Rao and Avinash~C Kak.
\newblock Retrieval from software libraries for bug localization: a comparative
  study of generic and composite text models.
\newblock In \emph{8th Working Conference on Mining Software Repositories},
  page 43–52, 2011.

\bibitem[Rehurek and Sojka(2010)]{rehurek2010software}
Radim Rehurek and Petr Sojka.
\newblock Software framework for topic modelling with large corpora.
\newblock In \emph{In Proceedings of the LREC 2010 Workshop on New Challenges
  for NLP Frameworks}. Citeseer, 2010.

\bibitem[Robillard et~al.(2012)Robillard, Bodden, Kawrykow, Mezini, and
  Ratchford]{Robillard-APIProperty-IEEETSE2012}
Martin~P. Robillard, Eric Bodden, David Kawrykow, Mira Mezini, and Tristan
  Ratchford.
\newblock Automated {API} property inference techniques.
\newblock \emph{IEEE Transactions on Software Engineering}, page~28, 2012.

\bibitem[R{\"o}der et~al.(2015)R{\"o}der, Both, and
  Hinneburg]{roder2015exploring}
Michael R{\"o}der, Andreas Both, and Alexander Hinneburg.
\newblock Exploring the space of topic coherence measures.
\newblock In \emph{Proceedings of the eighth ACM international conference on
  Web search and data mining}, pages 399--408, 2015.

\bibitem[Rosen and Shihab(2016)]{rosen2016mobile}
Christoffer Rosen and Emad Shihab.
\newblock What are mobile developers asking about? a large scale study using
  stack overflow.
\newblock \emph{Empirical Software Engineering}, 21\penalty0 (3):\penalty0
  1192--1223, 2016.

\bibitem[Rymer et~al.(2019)Rymer, Koplowitz, and Leaders]{rymer2019forrester}
John~R Rymer, Rob Koplowitz, and Salesforce~Are Leaders.
\newblock The forrester wave(tm) low-code development platforms for ad\&d
  professionals, q1 2019.
\newblock 2019.

\bibitem[Sahay et~al.(2020)Sahay, Indamutsa, Di~Ruscio, and
  Pierantonio]{sahay2020supporting}
Apurvanand Sahay, Arsene Indamutsa, Davide Di~Ruscio, and Alfonso Pierantonio.
\newblock Supporting the understanding and comparison of low-code development
  platforms.
\newblock In \emph{2020 46th Euromicro Conference on Software Engineering and
  Advanced Applications (SEAA)}, pages 171--178. IEEE, 2020.

\bibitem[salesforce()]{salesforce}
salesforce.
\newblock {Salesforce platform overview}.
\newblock {Available: \url{https://www.salesforce.com/in/?ir=1}}.
\newblock [Online; accessed 5-January-2021].

\bibitem[Sun et~al.(2015{\natexlab{a}})Sun, Li, Li, and
  Chen]{Sun-SoftwareMaintenanceHistoryTopic-CIS2015}
Xiaobing Sun, Bin Li, Yun Li, and Ying Chen.
\newblock What information in software historical repositories do we need to
  support software maintenance tasks? an approach based on topic model.
\newblock \emph{Computer and Information Science}, pages 22--37,
  2015{\natexlab{a}}.

\bibitem[Sun et~al.(2015{\natexlab{b}})Sun, Li, Leung, Li, and
  Li]{Sun-SoftwareMaintenanceTopic-IST2015}
Xiaobing Sun, Bixin Li, Hareton Leung, Bin Li, and Yun Li.
\newblock Msr4sm: Using topic models to effectively mining software
  repositories for software maintenance tasks.
\newblock \emph{Information and Software Technology}, 66:\penalty0 671--694,
  2015{\natexlab{b}}.

\bibitem[Sun et~al.(2016)Sun, Liu, Li, Duan, Yang, and
  Hu]{Sun-ExploreTopicModelSurvey-SNPD2016}
Xiaobing Sun, Xiangyue Liu, Bin Li, Yucong Duan, Hui Yang, and Jiajun Hu.
\newblock Exploring topic models in software engineering data analysis: A
  survey.
\newblock In \emph{17th IEEE/ACIS International Conference on Software
  Engineering, Artificial Intelligence, Networking and Parallel/Distributed
  Computing}, pages 357--362, 2016.

\bibitem[Thomas et~al.(2011)Thomas, Adams, Hassan, and
  Blostein]{Thomas-EvolutionSourceCodeHistoryTopic-MSR2011}
Stephen~W. Thomas, Bram Adams, Ahmed~E Hassan, and Dorothea Blostein.
\newblock Modeling the evolution of topics in source code histories.
\newblock In \emph{8th working conference on mining software repositories},
  pages 173--182, 2011.

\bibitem[Thomas et~al.(2014)Thomas, Adams, Hassan, and
  Blostein]{Thomas-SoftwareEvolutionUsingTopic-SCP2014}
Stephen~W. Thomas, Bram Adams, Ahmed~E Hassan, and Dorothea Blostein.
\newblock Studying software evolution using topic models.
\newblock \emph{Science of Computer Programming}, 80\penalty0 (B):\penalty0
  457--479, 2014.

\bibitem[Tian et~al.(2009)Tian, Revelle, and
  Poshyvanyk]{Tian-SoftwareCategorizeTopic-MSR2009}
Kai Tian, Meghan Revelle, and Denys Poshyvanyk.
\newblock Using latent dirichlet allocation for automatic categorization of
  software.
\newblock In \emph{6th international working conference on mining software
  repositories}, pages 163--166, 2009.

\bibitem[Uddin and Khomh(2017)]{uddin2017automatic}
Gias Uddin and Foutse Khomh.
\newblock Automatic summarization of api reviews.
\newblock In \emph{2017 32nd IEEE/ACM International Conference on Automated
  Software Engineering (ASE)}, pages 159--170. IEEE, 2017.

\bibitem[Vincent et~al.(2019)Vincent, Lijima, Driver, Wong, and
  Natis]{vincent2019magic}
P~Vincent, K~Lijima, Mark Driver, Jason Wong, and Yefim Natis.
\newblock Magic quadrant for enterprise low-code application platforms.
\newblock \emph{Retrieved December}, 18:\penalty0 2019, 2019.

\bibitem[vinyl()]{vinyl}
vinyl.
\newblock {Vinyl platform overview}.
\newblock {Available: \url{https://zudy.com/}}.
\newblock [Online; accessed 5-January-2021].

\bibitem[Wan et~al.(2019)Wan, Xia, and Hassan]{wan2019discussed}
Zhiyuan Wan, Xin Xia, and Ahmed~E Hassan.
\newblock What is discussed about blockchain? a case study on the use of
  balanced lda and the reference architecture of a domain to capture online
  discussions about blockchain platforms across the stack exchange communities.
\newblock \emph{IEEE Transactions on Software Engineering}, 2019.

\bibitem[Waszkowski(2019)]{waszkowski2019low-automating}
Robert Waszkowski.
\newblock Low-code platform for automating business processes in manufacturing.
\newblock \emph{IFAC-PapersOnLine}, 52:\penalty0 376--381, 01 2019.
\newblock \doi{10.1016/j.ifacol.2019.10.060}.

\bibitem[Wong et~al.(2019)Wong, Driver, and Vincent]{wong2019low}
Jason Wong, Mark Driver, and Paul Vincent.
\newblock Low-code development technologies evaluation guide, 2019.

\bibitem[Yang et~al.(2016)Yang, Lo, Xia, Wan, and Sun]{yang2016security}
Xin-Li Yang, David Lo, Xin Xia, Zhi-Yuan Wan, and Jian-Ling Sun.
\newblock What security questions do developers ask? a large-scale study of
  stack overflow posts.
\newblock \emph{Journal of Computer Science and Technology}, 31\penalty0
  (5):\penalty0 910--924, 2016.

\bibitem[zohocreator()]{zohocreator}
zohocreator.
\newblock {Zoho Creator platform overview}.
\newblock {Available: \url{https://www.zoho.com/creator/}}.
\newblock [Online; accessed 5-January-2021].

\end{thebibliography}
}

\end{document}